# *In vitro* binding energies capture Klf4 occupancy across the human genome

**Anne Schwager[1,⋆], Jonas Neipel[1,2,⋆], Yahor Savich[1,2,3,⋆], Douglas Diehl[1], Frank Jülicher[2,3,4,✉], Anthony A. Hyman[1,4,✉], and Stephan W. Grill[1,3,4,✉]**

[1]Max Planck Institute of Molecular Cell Biology and Genetics, 01307 Dresden, Germany; [2]Max Planck Institute for the Physics of Complex Systems, 01187 Dresden, Germany; [3]Center for Systems Biology Dresden, 01307 Dresden, Germany; [4]Cluster of Excellence Physics of Life, Technical University Dresden, 01062 Dresden, Germany; ⋆A.S., J.N., and Y.S. contributed equally to this work.

**Transcription factors (TFs) regulate gene expression by binding to specific genomic loci determined by DNA sequence. Their sequence specificity is commonly summarized by a consensus binding motif. However, eukaryotic genomes contain billions of low-affinity DNA sequences to which TFs associate with a sequence-dependent binding energy. We currently lack insight into how the genomic sequence defines this spectrum of binding energies and the resulting pattern of TF localization. Here, we set out to obtain a quantitative understanding of sequence-dependent TF binding to both motif and non-motif sequences. We achieve this by first pursuing accurate measurements of physical binding energies of the human TF Klf4 to a library of short DNA sequences in a fluorescence-anisotropy-based bulk competitive binding assay. Second, we show that the highly non-linear sequence dependence of Klf4 binding energies can be captured by combining a linear model of binding energies with an Ising model of the coupled recognition of nucleotides by a TF. We find that this statistical mechanics model parametrized by our *in vitro* measurements captures Klf4 binding patterns on individual long DNA molecules stretched in the optical tweezer, and is predictive for Klf4 occupancy across the entire human genome without additional fit parameters.**

**Correspondence:** ***hyman@mpi-cbg.de, julicher@pks.mpg.de, grill@mpi-cbg.de***

**ACKNOWLEDGEMENTS**

We thank Sina Wittmann for extensive help at the beginning of this project. We thank Andrei Pozniakovsky (MPI-CBG) for providing the DNA construct for protein expression. We thank Titus Franzmann (Dresden University of Technology) for valuable advice and discussions. We acknowledge Anupam Singh, Kshitij Deshpande, Theresia Gutmann, Jie Wang, and David Kuster (MPI-CBG) for helpful discussions related to the project. We are grateful for the technical support provided by Barbara Borgonovo, Régis Lemaitre, Eric Geertsma, and Aliona Bogdanova (Protein Biochemistry Facility, MPI-CBG), as well as Martin Stöter, Nadine Tomschke, Antje Janosch, and Rico Barsacci (Technology Development Studio, MPI-CBG), and Michael Bugiel (Biophysical Core Facility, MPI-CBG). We acknowledge support from the NOMIS Foundation (A.A.H.), the Volkswagen Foundation (F.J. and A.A.H.), and the Max Planck Society.

**AUTHOR CONTRIBUTIONS**

A.S. and Y.S. pursued experimental design with input from F.J., A.A.H., and S.W.G.. A.S. performed the experiments, and D.D. performed the optical tweezer experiments. A.S., J.N., Y.S., and D.D. analyzed the data, with input from all authors. J.N. and Y.S. developed the theory with input from all authors. A.S., J.N., Y.S., and S.W.G. wrote the manuscript with input from all authors.

Regulating which, when and how genes are expressed is what allows biological cells to exist, function, and respond to their environment. Transcription factors are the central component of the regulation of gene expression (Jacob and Monod, 1961). TFs are proteins that bind to DNA to regulate which sequences are transcribed. In order to regulate a certain gene, a TF needs to bind at the right place in the genome (Lambert et al., 2018), picking out a few locations over a vast number of possible locations (von Hippel and Berg, 1986; Berg and von Hippel, 1987). TFs can achieve this because they, when bound to DNA, directly interact with a specific number of DNA basepairs in a DNA sequence-dependent manner, thus yielding a sequence specific binding energy (Maerkl and Quake, 2007; Tjian and Maniatis, 1994; Kadonaga, 2004). The genomic locations to which TFs associate with high affinity are well studied (Ovek Baydar et al., 2025; Weirauch et al., 2014; Lambert et al., 2018), but we still lack a quantitative understanding of the sequence-dependence of the TF-DNA binding energy, accurate across the entire spectrum of possible sequences (Mahony and Pugh, 2015; Yan et al., 2021; Shahein et al., 2022; Kribelbauer et al., 2019).

Here we focus on the human transcription factor Krüppel-like factor 4 (Klf4), one of the four Yamanaka factors that are sufficient to experimentally reprogram somatic cells into induced pluripotent stem cells (An et al., 2019; Nishimura et al., 2014). As a pioneer TF, Klf4 is capable of binding across the human genome, also in regions of compacted chromatin (Soufi et al., 2012). This allows Klf4 to bind thousands of genomic regulatory elements, including promoters and enhancers, to directly regulate hundreds to thousands of genes in a cell-type-dependent manner (Moonen et al., 2022; Lambert et al., 2018). We here set out to investigate how DNA sequence determines where and how strongly Klf4 binds across the human genome.

TF binding to genomic DNA is typically quantified by Chromatin Immunoprecipitation sequencing (ChIP-seq), i.e. by crosslinking TFs to the bound DNA sequence in chemically fixated cells, followed by sequencing of the crosslinked DNA (Johnson et al., 2007). ChIP-seq allows to identify 100-1000bp wide regions in the genome to which a TF is stably bound (Landt et al., 2012). This allows for the identification of those DNA sequences to which the TF preferentially binds. These are captured by a consensus sequence motif (Lewin, 1974; Schmidt et al., 2009; Johnson et al., 2007; Inukai et al., 2017). TF-DNA binding can also be characterized *in vitro*, using Systematic Evolution of Ligands by EXponential enrichment (SELEX). This is a high-throughput *in vitro* method that identifies sequence patterns that optimize TF binding, starting from a large pool of short random sequences (Jolma et al., 2013, 2010).

However, TF binding is not limited to a small set of optimal sequences. In fact, more recent evidence suggests that TF binding to low affinity sequences enables overall high sequence specificity (Shahein et al., 2022; Kribelbauer et al., 2019). In addition, the evolution of complex gene regulation in eukaryotes has been accompanied by a reduction of motif length as compared to prokaryotes, which reduces the capability to reliably identify specific genomic locations (Wunderlich and Mirny, 2009). This apparent reduction in specificity of TF binding is paradoxical, given the significant increase in genome size, and the increased complexity of gene regulation in eukaryotes as compared to prokaryotes. Importantly, a common feature of eukaryotic TFs is that they contain long intrinsically disordered regions (IDRs) that enable multivalent interactions (Minezaki et al., 2006; Brodsky et al., 2021; Abidi et al., 2025). All this implies that the sequence specificity of eukaryotic TFs results from a complex multitude of individually weak interactions. Thus, to understand the regulatory complexity encoded in eukaryotic genomes, we need to obtain a quantitative understanding of the binding of TFs together with their IDRs to both the strongly binding and all the weakly binding DNA sequences (Már et al., 2023).

Crucially, *in vitro* studies of TF binding are typically pursued with protein fragments that do not include the extensive IDRs. To the best of our knowledge, not a single *in vitro* study has assessed the sequence specificity of full-length Klf4 (Sun et al., 2013; Schaepe et al., 2025). Moreover, most studies have focused on identifying sequence motifs, in contrast to measuring the sequence-dependent binding energy (Inukai et al., 2017). While a sequence motif allows to identify high-affinity binding sites, we here pursue a statistical mechanics approach based on the direct measurement of sequence-dependent binding energies to shed light on the statistics of Klf4 binding across the genome.

The strength of transcription factor binding to a particular DNA sequence is quantified by a binding energy $\Delta G$, where lower values indicate stronger binding. Relevant for sequence-dependent TF binding is the binding energy difference $\Delta\Delta G$ between different sequences that compete for binding. In the first part of the paper we measure the binding energy difference $\Delta\Delta G$ with respect to a reference sequence across a library of 73 designed sequences. To measure $\Delta\Delta G$ with sub-$k_BT$ accuracy we pursue Fluorescence Anisotropy (FA) experiments in a competitive binding mode (Roehrl et al., 2004). The binding strength of a TF to a particular DNA sequence is typically quantified in terms of the dissociation constant $K_d$, defined as the ratio of the product of equilibrium concentrations of free TF and free DNA over the concentration of the TF–DNA complex. This corresponds to the TF concentration at which half of DNA molecules are bound at equilibrium. Here we pursue competitive binding experiments that directly measure the ratio $K_d/K_{d,\mathrm{ref}}$ of the dissociation constants of the two sequences that are present, and thus their binding energy difference $\Delta\Delta G = k_BT \log K_d/K_{d,\mathrm{ref}}$.

The space of possible DNA sequences is astronomically large, even short motifs of length L span $4^L$ combinations, so direct experimental measurement of binding for every sequence is impossible. To generalize from limited data, models of transcription factor sequence specificity have been developed to predict the binding across sequence space (Bussemaker et al., 2001). The canonical model is the position weight matrix (PWM), an additive (linear) per-position model that originated from the biophysical energy-matrix view of Berg and von Hippel (von Hippel and Berg, 1986). It remains widely used because it focuses on the consensus motif, and is compact and interpretable. PWMs (and related motif models) are excellent at ranking candidate sites and locating likely binding positions in genomes, but their underlying independence and linearity assumptions mean the numeric PWM score is not a reliable proxy for binding energy or quantitative occupancy in many cases (Maerkl and Quake, 2007). This shortcoming has motivated richer sequence models, i.e. higher-order PWMs, Hidden Markov Models and modern machine-learning/deep-learning approaches, that improve site prediction by capturing dependencies and complex syntax (Zhao et al., 2012; Mathelier and Wasserman, 2013; Avsec et al., 2021; Siddharthan, 2010; Omidi et al., 2017). Recent high-throughput assays that infer relative binding energies across many thousands–to–millions of sequences make the latter approach increasingly practical, enabling models that can be trained and validated on energetic quantities (Le et al., 2018; Maerkl and Quake, 2007).

In the second part of this paper we put forward a statistical mechanics model of sequence-dependent relative energy $\Delta\Delta G$ of Klf4 binding to DNA. This then allows us to go beyond site predictions and provide occupancy statistics of Klf4 on extended stretches of DNA, including the entire human genome. For this we consider an Ising model of cooperative nucleotide recognition by Klf4. In this equilibrium model, a sequence-dependent energy defines the statistics of recognition states internal to Klf4, which gives rise to a non-linear sequence dependence of the binding free energy. With this, we can then capture the competitive binding assay where many Klf4 molecules individually bind to short DNA oligonucleotides (oligos) of two different sequence types, as well as the measured occupancy of many Klf4 molecules as a function of position along a single extended DNA molecule stretched in an opti-cal tweezer. Strikingly, this equilibrium model of Klf4 binding is also capable of predicting Klf4 occupancy frequencies across the entire human genome.

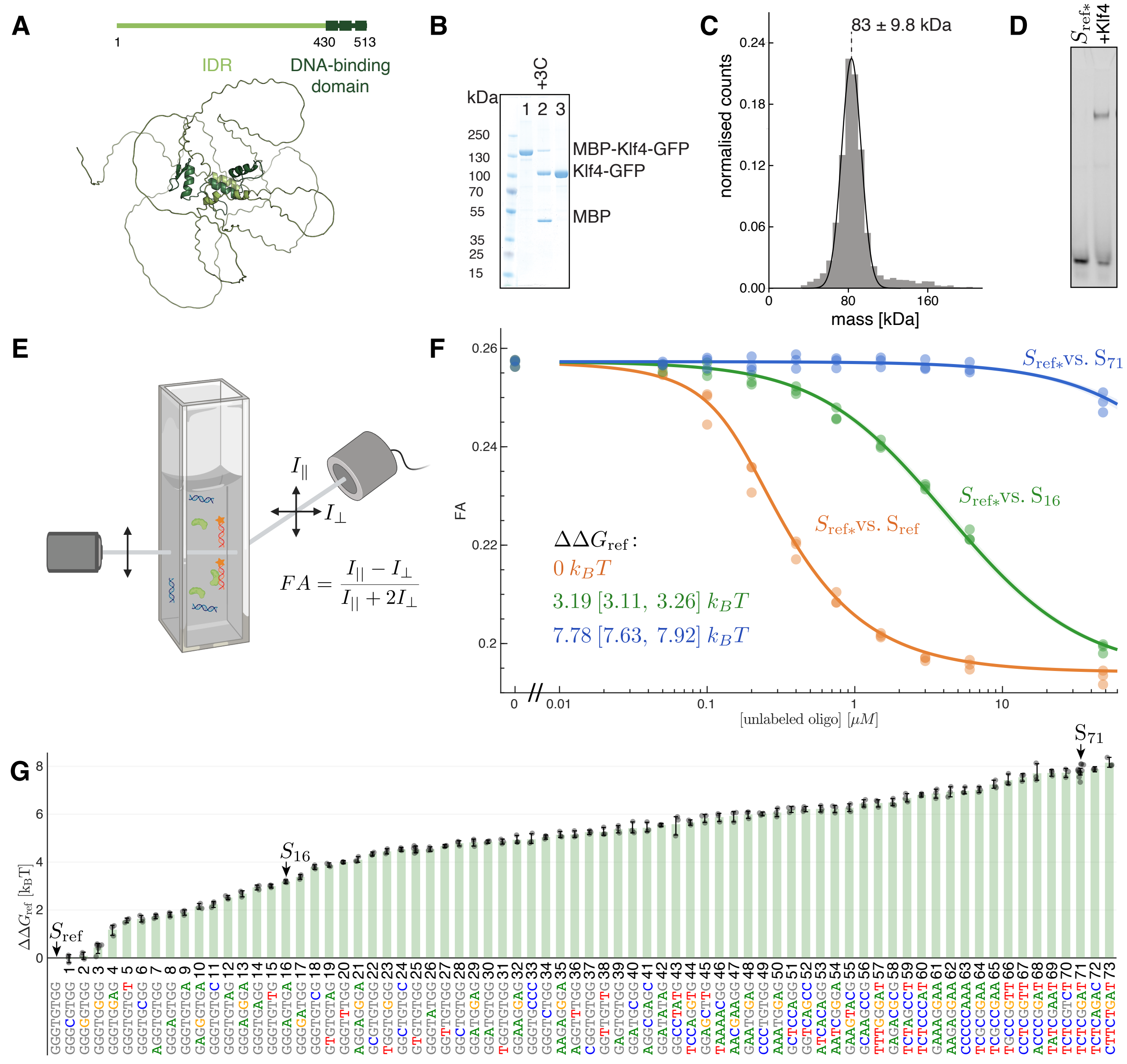

**Figure 1. Measurement of Klf4-DNA binding energies (A)** Top, schematic of the transcription factor Klf4 showing the N-terminal intrinsically disordered region (IDR, 430 aa, light green) and the C-terminal DNA-binding domain (83 aa, dark green) containing three C2H2 zinc fingers. Bottom, AlphaFold model of human Klf4 (UniProt: O43474-3, rendered in PyMol). **(B)** Coomassie-stained SDS-PAGE gel of purified Klf4-GFP. Line 1, N-terminal MBP and C-terminal GFP. Line 2, cleavage with 3C. Line 3, removed MBP, yielding Klf4-GFP (83 kDa). **(C)** Mass photometry performed on Klf4-GFP (hereafter Klf4) reveals a histogram of measured molecular mass, bar width is $6.42\,kDa$ (curve; Gaussian fit with mean $\sim 83\,kDa$ and standard deviation $9.8\,kDa$). **(D)** EMSA reveals an upshift upon addition of Klf4 to the strong-binding oligonucleotide (oligo) $S_{\mathrm{ref}}$ labeled with Cy5. **(E)** Schematic of the fluorescence anisotropy assay. Linearly polarized excitation light passes through a mixture of bound and unbound fluorescent oligos, emission is detected in parallel and perpendicular orientations. Anisotropy decreases upon ATTO550-labeled oligos unbinding from Klf4. **(F)** Competition FA assay; Klf4 and ATTO550-labeled oligo $S_{\mathrm{ref}*}$ is competed out by three different unlabeled oligos: $S_{\mathrm{ref}}$ in orange, $S_{16}$ in green, $S_{71}$ in blue. At high concentrations the strong competitor $S_{\mathrm{ref}}$ fully displaces the labeled oligo $S_{\mathrm{ref}*}$. The $S_{16}$ oligo (green) shows intermediate competition, and the weak competitor ($S_{71}$, blue) shows minimal displacement. **(G)** Measured binding energy $\Delta\Delta G_{\mathrm{ref}}$ for all 73 oligos, ordered by the mean. Bottom, the central 8-bp sequences are shown for each oligo, where nucleotides in color are changed with respect to $S_{\mathrm{ref}}$. Sequences used in **G** are indicated by arrows. Bars show the mean of individual experiments; black dots mark individual measurements, horizontally offset for clarity; error bars indicate the 95% confidence interval for the mean as determined by bootstrapping.

## Results

### Quantification of Klf4 binding energies *in vitro*

To facilitate fluorescence-based occupancy measurements we purified full length Klf4-GFP (referred to in the following as Klf4; see Fig. 1A), optimized from previous protocols (Morin et al., 2022). We affinity-purified full-length Klf4 from insect cells using the baculovirus expression system (Fig. 1B). The purified Klf4-GFP appeared as a monomer, as demonstrated by mass photometry (Fig. 1C) and size-exclusion chromatography coupled with static light scattering (Fig. S1). We confirmed that the protein was free of host nucleic acid contamination using UV spectrophotometry (Methods), and we confirm functional DNA–protein interaction by conducting an Electrophoretic Mobility Shift Assay (EMSA; Fig. 1D).

First, we assayed binding of Klf4 to short oligos where residual ssDNA after annealing was removed with HPLC. We evaluated Klf4 binding to a 17 basepair long fluorescently labeled DNA oligo that contains the canonical binding motif of Klf4 (Chen et al., 2008), confirming first-order equilibrium binding kinetics (Fig. S1C,D). We next measure $\Delta\Delta G$ via bulk competitive binding, where a second unlabelled DNA oligomer competes for Klf4 binding. As the unlabelled oligo competes with the labelled one for Klf4 binding, the FA signal decreases. $\Delta\Delta G$ is then determined via a single-parameter fit of the titration curve of the unlabelled oligo (Fig. 1F, SI).

We measured 73 unbinding curves in FA competition experiments, which resulted in 73 measurements of $\Delta\Delta G$ with an average uncertainty $\lesssim 0.2\,k_BT$, revealing a sequence-dependent spectrum of Klf4 binding energy differences $\Delta\Delta G$ that reach up to $\sim 8\,k_BT$ (Fig.1G).

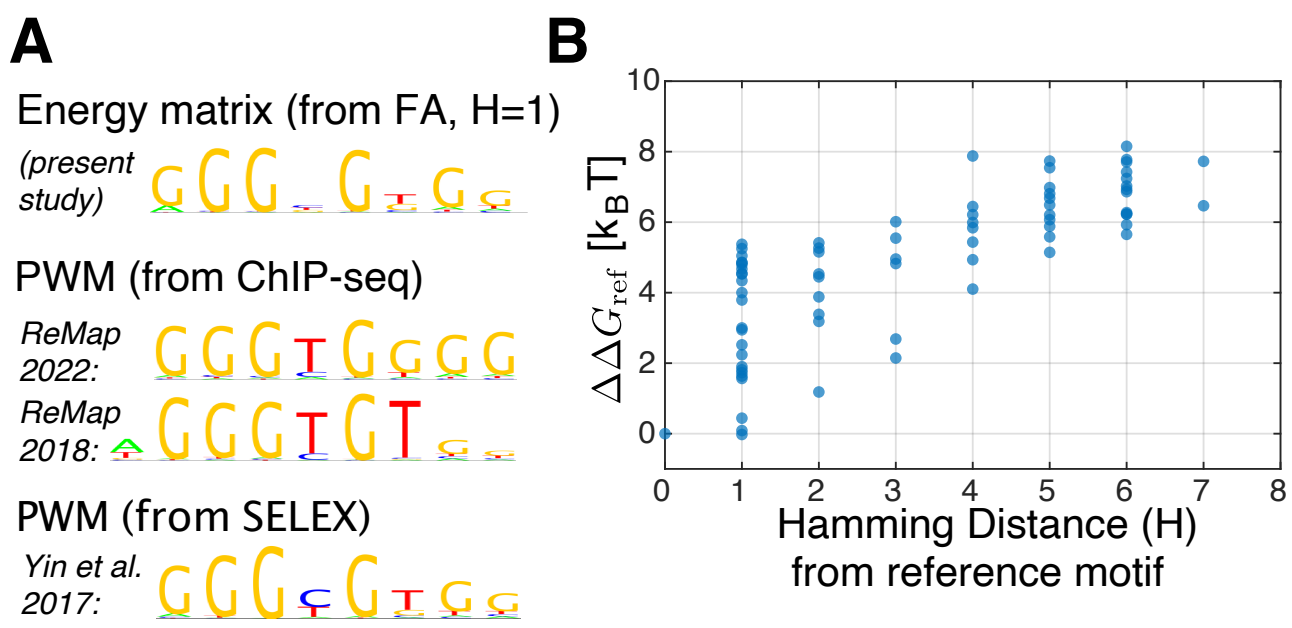

**Figure 2. Klf4 binds a poly-G/C motif with high specificity. (A)** Motif logos for linear single-nucleotide models of the sequence preference of Klf4. Letter sizes represent the information content of the corresponding entry in the energy or weight matrices. We compare the energy matrix obtained from energy measurements (Fig. 1G) of $H = 1$ sequences (upper logo) to position-weight matrices to the literature (lower logos). The ChIP-seq PWMs were constructed from cells from human cell lines from various studies aggregated by ReMap (Hammal et al., 2021; Chèneby et al., 2017) and retrieved as a versioned motif from the JASPAR database (MA0039.5 and MA0039.3) (Ovek Baydar et al., 2025) The SELEX-derived PWM was obtained from the CIS-BP database (M04667_3.00) from *in vitro* data obtained in (Yin et al., 2017). **(B)** Measured relative binding energies $\Delta\Delta G_{\rm ref}$ (Fig. 1G) plotted as a function of the distance $H$ from the reference sequence.

## Linear energy models

We next set out to identify a sequence-dependent binding energy model that accurately captures the measured values of $\Delta\Delta G$ for all the sequences probed. A basic feature of any sequence is the GC content (Liu et al., 2014; Wang et al., 2012; Tillo and Hughes, 2009; Duret and Galtier, 2009). We first consider an energy model linear in GC content, but we find this to result in poor predictions for the measured $\Delta\Delta G$ of all sequences probed, with a root mean square difference of 1.7 $k_BT$ (Fig. S9D). We next consider an additive linear model where each nucleotide of the sequence bound to the TF contributes independently to $\Delta\Delta G$. This energy model is parameterized with an Energy Matrix closely related to PWMs (Bussemaker et al., 2001) used to characterize sequence motifs (see SI). Such a linear single nucleotide model can capture the sequence dependence of $\Delta\Delta G$ for sequences in the immediate vicinity of a reference motif, i.e. for a Hamming distance $H = 1$ corresponding to single substitutions (Bintu et al., 2005; Djordjevic et al., 2003; Bussemaker et al., 2001). We utilized $\Delta\Delta G$ measurements of all $H = 1$ sequences to identify the energy matrix that exactly captures binding energy difference in the vicinity of the reference motif $S_{\rm ref}$ (see Fig. 2A). In Fig. 2A we illustrate this energy matrix as a motif logo. We note that the identified logo compares favorably to those determined via ChIP-seq of human cell lines (Hammal et al., 2021; Chèneby et al., 2017), or via SELEX (Yin et al., 2017).

Fig. 3A shows that while the substitution of a first nucleotide results in an increase of the binding energy difference $\Delta\Delta G(S)$ by a value of up to $5\,k_BT$, the binding energy difference saturates with more substitutions around $8\,k_BT$ (Fig. 3B). A linear model cannot capture such a saturation behavior (Maerkl and Quake, 2007). Hence, we find that the linear model significantly overestimates $\Delta\Delta G$ for non-motif sequences with Hamming distance of $H \geq 2$. In Fig. S10D, we obtained an energy matrix by fitting to all measured sequences with $H \leq 3$. While this model allows to capture binding to $H \leq 3$ sequences (RMSD: $0.7 k_BT$), it essentially fails for $H > 3$ (RMSD: $2.8 k_BT$). This again indicates that a linear model cannot accurately capture $\Delta\Delta G$ across sequence space. In what follows we put forward a statistical mechanics model of TF binding that accurately captures the binding to any sequence.

## Ising model of sequence recognition

Previous studies have considered that TF's can switch between a weakly bound state referred to as a 'test state' (Horton et al., 2023), and a strongly bound state. We here put forward a statistical mechanics model of sequence recognition where the regions of the TF that interface to the nucleotides probed can each switch between two states in a cooperative manner. In this Ising model, $\sigma = +1$ refers to the state where a nucleotide is recognized and strongly bound, and $\sigma = -1$ if the nucleotide is not recognized and thus bound to Klf4 in a sequence-independent alternative mode. The energy of the strongly bound state with respect to the alternative binding state is given by the position and nucleotide dependent energy $\Delta\epsilon_i$. For simplicity, we consider TF binding in the alternative mode as sequence-independent. Crucially, recognition of neighboring nucleotides is coupled in the spirit of an Ising model by the coupling constant $J$.

The overall binding energy $\Delta\Delta G$ to a sequence is given by the logarithm of the partition function (see SI). Importantly, although the microscopic energy of a microstate state is linear in the nucleotide-specific contributions $\Delta\epsilon_i$ in our cooperative Ising model, $\Delta\Delta G$ is non-linear in $\Delta\epsilon_i$. This nonlinearity arises because the total binding energy is obtained by summing over all configura-tional microstates of the TF-DNA complex in the partition function. Due to the cooperative Ising-type coupling, unfavorable nucleotides (i.e. $\Delta\epsilon_i > 0$) locally destabilize the strongly bound configuration and thereby reduce the probability $p(\sigma_j > 0)$ that neighboring positions $j$ are simultaneously recognized. As a consequence, once one or a few nucleotides are unfavorable, the TF is already biased toward partially or fully unrecognized configurations. Introducing additional unfavorable nucleotides therefore has a progressively smaller effect on the overall binding energy. This leads to a saturating dependence of $\Delta\Delta G$ on sequence quality: binding energies decrease approximately linearly for weak perturbations around the optimal sequence (i.e. $H = 1$), but flatten for increasingly unfavorable sequences (i.e. $H > 1$) as cooperative recognition becomes unlikely.

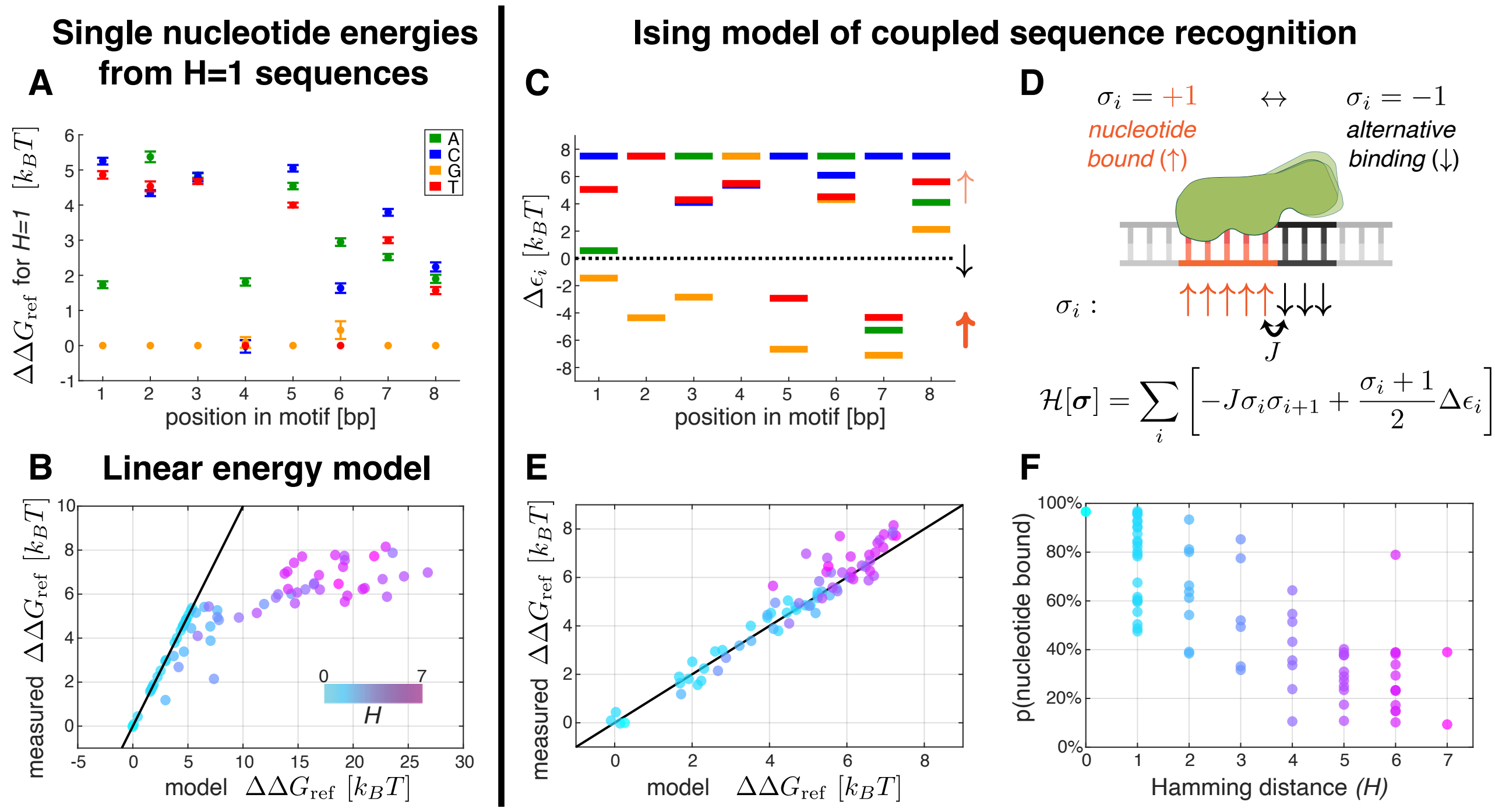

**Figure 3. Ising model of sequence recognition captures the non-linear sequence dependence of Klf4 binding energies (A)** Measured change in Klf4 binding energy $\Delta\Delta G_{\mathrm{ref}}$ upon single nucleotide substitutions with respect to the reference motif. Data points represent the measured $\Delta\Delta G_{\mathrm{ref}}$ for $H = 1$ sequences where the nucleotide at the given position in the motif ($x$-axis) has been replaced by the nucleotide indicated by the color. Error bars indicate the $95\%$ confidence interval from bootstrapping. Points without error bars represent the reference sequence with $\Delta\Delta G_{\mathrm{ref}} = 0$. **(B)** Prediction of binding energies for all library sequences from the energy matrix obtained from the measured $\Delta\Delta G_{\mathrm{ref}}$ of the $H = 1$ sequences in **A**. Color of data points indicates Hamming distance $H$ with respect to reference motif with sequences close to reference in cyan. Black line indicates $x = y$. **(C)** Nucleotide binding energies $\Delta\epsilon_i$ (colored bars) of our Ising model obtained by fitting to the measured $\Delta\Delta G_{\mathrm{ref}}$ for $H \leq 3$. $\Delta\epsilon_i$ is the energy of the *nucleotide bound* with respect to the *alternative binding* state with negative $\Delta\epsilon_i$ indicating energetically favorable nucleotide binding. Note that we specify an upper bound $\max\Delta\epsilon_i = 7.5k_BT$ in order to obtain a robust fit (see SI). Color of bars indicates nucleotide identity as in **A**. **(D)** Illustration of statistical mechanics model explained in main text, defined by the Hamiltonian $\mathcal{H}$ of single nucleotide states $\sigma_i = \pm 1$. Schematic shows TF (green) bound to DNA, where some nucleotides are strongly bound in a sequence-dependent manner (red, upward arrows corresponding to $\sigma_i = +1$) whereas the remainder of the bound sequence is bound in the alternative mode (dark grey, $\sigma_i = -1$). The binding states of neighboring nucleotides is coupled in terms of the coupling constant $J$. For the fit in **C,E**, we use $J = 2k_BT$. **(E)** Comparison of predicted to measured relative binding energies as in **B**, but using the Ising model of nucleotide recognition (**C,D**). Note that we take here also into account binding to the reverse strand and flanking sequences in contrast to **B**. In Fig. S10, we consider also a linear energy where binding to flanking sequences is taken into account, analogously to **E**. **(F)** Average fraction of *nucleotides bound* out of the TF-bound 8bp sequence , i.e. $p(\sigma_i = +1)$ calculated with the Ising model for the library sequences.

We set out to test whether such a model can capture the measured Klf4 binding energies to 17bp oligos. We calculate the binding energy $\Delta\Delta G$ to a 17 bp oligo by calculating the partition function taking into account binding to all 8-mers on the two strands of the oligos, (Horton et al., 2023). We obtain the nucleotide-specific binding energies $\Delta\epsilon_i$ by fitting this Ising model to our bulk data (Fig. 1G) for sequences with $H \leq 3$ (see SI, Fig. 3C). By optimizing the prediction for all sequences (see SI), we obtain $J \sim 2\,k_BT$. This corresponds to a regime where nucleotide recognition is strongly coupled. When using this model, calibrated by the $H \leq 3$ sequences, to predict $\Delta\Delta G$ for the remaining $H > 3$ sequences, we find that it predicts $\Delta\Delta G$ to within $0.8\,k_BT$ (Fig. 3E). We find that for non-motif sequences, i.e. $H > 3$, $\sim 30\%$ of nucleotide interfaces remain in the strongly bound state (Fig. 3D). We conclude that a statistical mechanics model of equilibrium binding that takes into account coupled recognition of nucleotides in an Ising sense reproduces the saturation behaviour and accurately captures binding energies measured in bulk competition experiments for sequences both close to and far away from the reference sequence (Figs. 2B and 3B,E).

### Klf4 occupancy patterns on long stretches of DNA

Our statistical mechanics model yields predictions of $\Delta\Delta G(S)$ of the relative Klf4 binding energy for any sequence $S$ based on direct *in vitro* measurements of $\Delta\Delta G$ using short oligonucleotides. We next asked if this model is capable of capturing the occupancy of Klf4 on a single long piece of DNA, where many sequences simultaneously compete for binding. To this end we stretch a single piece of $\lambda$-DNA (48.5 kbp) to full extension in an optical tweezer (to $\sim 15\,\mathrm{pN}$ of tension), and subject it to a dilute solution of our GFP-labelled Klf4 (12 - 70 nM). These concentrations are below the prewetting concentration of Klf4, and Klf4 molecules are expected to individually associate with DNA (Morin et al., 2022). We note that, in comparison to (Morin et al., 2022), Klf4 was here purified with an improved protocol (see Methods). We monitor binding of GFP-labelled Klf4 to the single molecule of $\lambda$-DNA using confocal microscopy (Fig. 4A). Since we can quantify Klf4 localization on DNA with a resolution of about $\sim 1\mathrm{kbp}$ or $\sim 300\,nm$, this allows us to translate spatial localization of Klf4 on DNA to sequence localization.

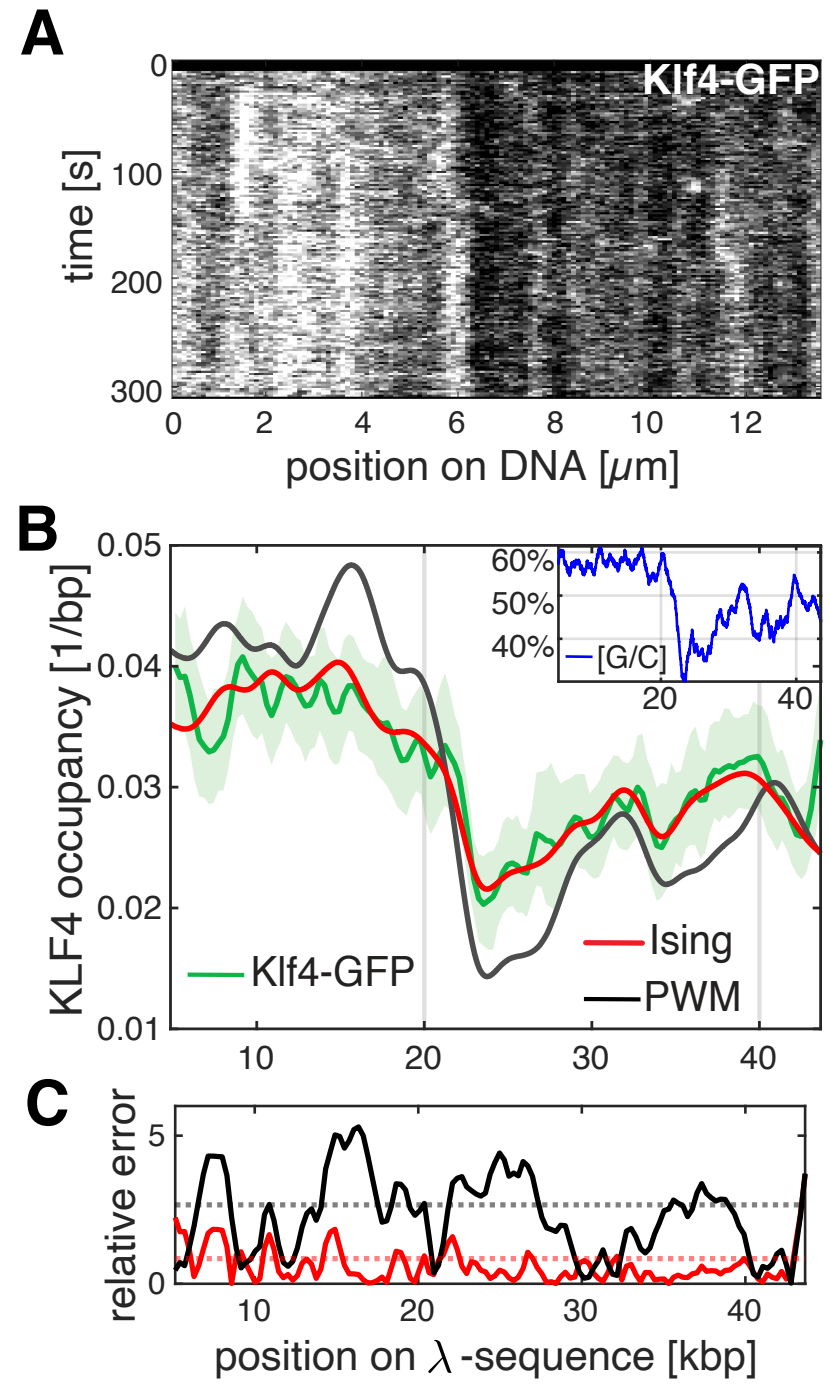

**Figure 4. Ising model predicts Klf4 occupancy pattern on kbp DNA molecules (A)** Kymograph showing fluorescent intensity of Klf4-GFP along an individual $\lambda-$ DNA stretched in the optical tweezer as a function of time post exposure to a solution containing $38\,nM$ of Klf4. **(B)** Klf4 occupancy (molecules per bp) along $\lambda$-DNA. Green curve shows the average measured Klf4 occupancy obtained from optical tweezer experiments as in **A** ($n = 57$ with varying bulk concentration between $12.5\,nM$ and $75\,nM$). Shaded area indicates $95\%$ confidence interval of the mean from bootstrapping. Red and black curves show model predictions for Klf4 occupancy using our Ising model (red curve, Fig. 3) and a linear model parametrized by the PWM from (Hammal et al., 2021) (black curve). Inset: GC content of $1\,\mathrm{kbp}$ intervals along the $\lambda$-sequence. **(C)** Relative prediction error, i.e. absolute difference in Klf4 occupancy between average measurement and the theory predictions, normalized with respect to the standard error of the mean of the measurement. Red and black curves indicate prediction errors of Ising model and linear energy model, respectively, as in **B**. Dashed lines show the root mean square of the respective relative prediction error.

At this resolution, the dominant statistic of any sequence is the GC content. We find that Klf4 preferentially binds to G/C-rich regions (Fig. 4B). Notably, we find that the Ising model parametrized by *in vitro* data predicts the measured Klf4 occupancy pattern to a quantitative degree and within the range of experimental uncertainty (Fig. 4B,C). Notably, our Ising model provides a predicted occupancy that is considerably more accurate than that of a linear energy model based on the PWM obtained from ChIP-seq (Chèneby et al., 2017; Hammal et al., 2021) (see black curve in Fig. 4B,C). We conclude that *in vitro* binding energies measured with a library of short DNA oligos, via the use of an Ising model, accurately predict the Klf4 occupancy pattern on a long and stretched piece of DNA.

### Genomic energy landscape of Klf4

We next asked if our equilibrium model can predict the occupancy pattern of Klf4 along the human genome. In order to facilitate a comparison to ChIP-seq data obtained in human cells (Chen et al., 2008), we determine the genomic binding energy landscape by calculating $\Delta\Delta G$ of Klf4 binding for overlapping 100bp intervals. We then bin all 100bp intervals of the human genome according to the calculated *in vitro* binding energy $\Delta\Delta G$ in percentiles. We then, for each energy bin, use the number of ChIP-seq peaks to estimate an *in vivo* Klf4 occupancy $p$ in units of a genome-wide average $p_0$ (see SI). In thermodynamic equilibrium, we would expect the relative occupancy $p/p_0$ to be given by the Boltzmann weight $\exp(-\Delta\Delta G/k_BT)$ associated with the relative binding energy $\Delta\Delta G$, as long as binding is dilute, i.e. $p \ll 1$. We thus expect a linear relationship with slope -1 when plotting the logarithm of measured occupancy versus binding energy, i.e. $\log(p/p_0) = -\Delta\Delta G/k_BT + c$, where $c$ is a constant. Fig. 5A reveals a remarkably close agreement with this prediction over a broad range of binding energies (for more than $\sim 4\,k_BT$). We note that linear energy models, parameterized either with the PWM obtained from ChIP-seq or from our measured energies of our $H = 1$ sequences (Fig. 1-3), fail to capture a linear relationship with slope -1 (Fig. 5B). We conclude that *in vitro* binding energies measured with a library of short DNA oligos, via the use of an Ising model, predict occupancy statistics of Klf4 along the entire genome in human cells. Together, this implies that 1) *in vitro* binding energies are quantitatively meaningful *in vivo*, and that 2) Klf4 binding in human cells can be captured by equilibrium physics despite the complex non-equilibrium nature of the cell nucleus.

## Discussion

Here, we obtained direct measurements of Klf4 binding energies to a library of short DNA oligos, achieving sub-$k_BT$ accuracy over a range of $\sim 8\,k_BT$, using a bulk competitive FA assay. Recently, a competitive FA assay with higher throughput has been used to quantify the sequence-dependent binding of other TFs (Jung et al., 2018) using protein fragments that contained the DNA-binding domain only. We here measured relative binding energies of full-length Klf4 in solution under near physiological buffer conditions. Notably, us-ing unlabeled oligo concentrations of up to $\sim 50\,\mu M$ allowed us to capture binding energy differences of up to $\sim 8\,k_BT$, corresponding to about 3000-fold difference in binding affinity as quantified by $K_d$. We note that removing residual ssDNA after annealing was essential to avoid ssDNA-related artifacts at the $\mu$M concentrations of the unlabeled oligos (Wysoczynski et al., 2013).

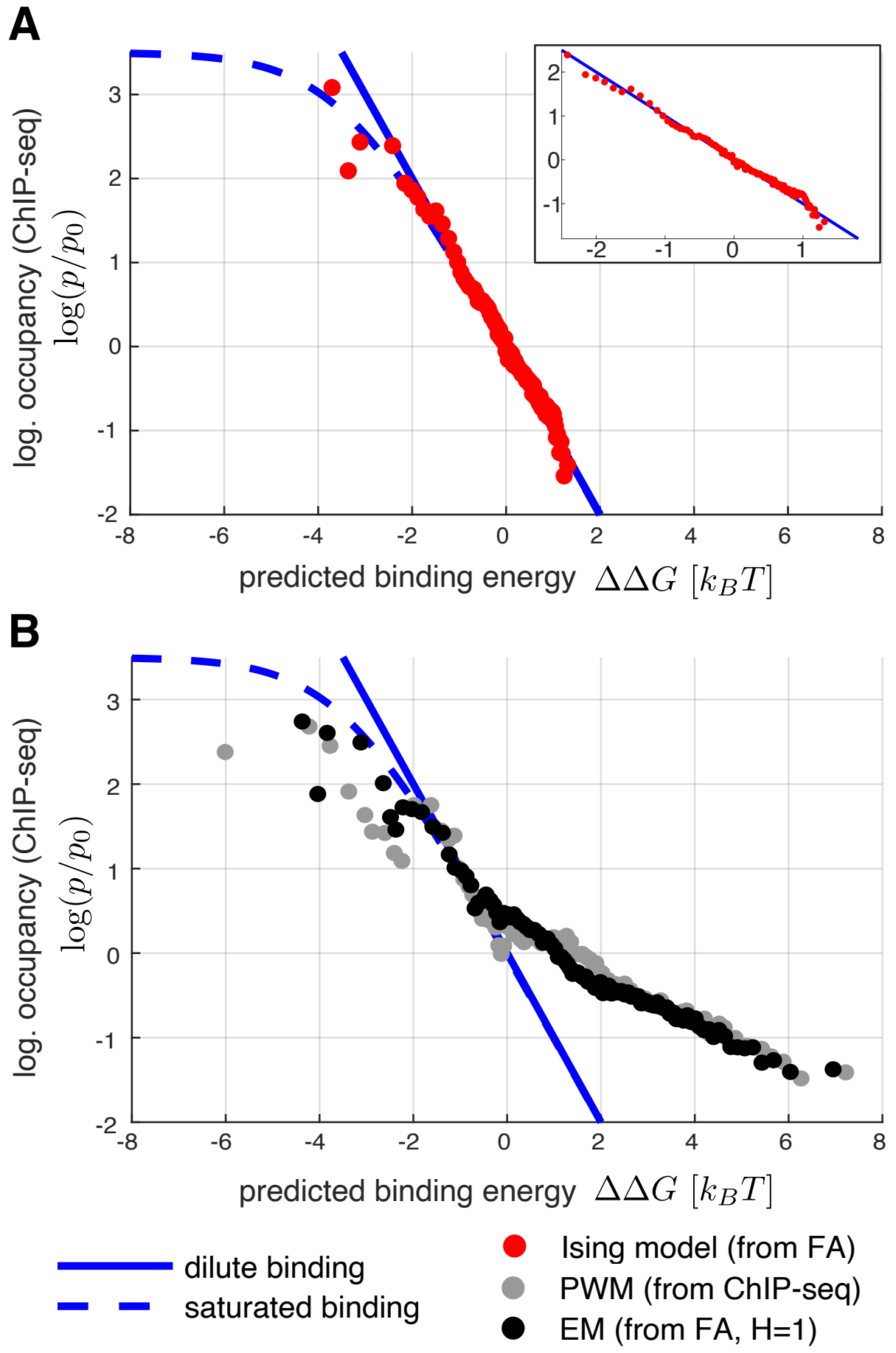

**Figure 5.** ***In vitro* binding energies captures occupancy statistics of Klf4 across the human genome.** **(A)** scatter plot of the Klf4 occupancy of energy percentiles. Each data point corresponds to a set of 100bp intervals of the human genome, corresponding to a percentile of the distribution of binding energies $\Delta\Delta G$ calculated with the Ising model (Fig. 3), using the same $J$ and $\Delta\epsilon_i$ as in Figs. 3 and 4 but using a temperature $T = 37°C$, the standard culture temperature for human cells, instead of the temperature $T = 21.5°C$ at which FA measurements were performed. Note that we define here $\Delta\Delta G$ with respect to the genome wide average binding energy instead of a specific reference sequence. $y$-axis shows Klf4 occupancy in terms of the logarithm of the fraction $p$ of Klf4-occupied sequences in an energy percentile, normalized with respect to the genome-wide average occupancy $p_0$. Here occupied sequences are identified as the 100bp intervals surrounding a ChIP-seq peak in a data set from human cells (Hammal et al., 2021). Blue lines indicate the predicted occupancy as a function of binding energy considering equilibrium binding kinetics of Klf4. Blue solid line ($y = -x$) corresponds to dilute binding $p \ll 1$. Blue dashed line indicates saturated binding, i.e. the predicted occupancy for a chemical potential $\mu = -3.5\, k_BT$ such that $\sim 2\%$ of the considered sequences are on average occupied. **(B)** Klf4 occupancy statistics as in **A** but with binding energy predicted by a linear energy model parametrized by either the PWM inferred from the same ChIP-seq data (black dots) or the energy matrix (EM) from FA energy measurements of $H = 1$ sequences (gray dots).

We next developed an Ising model of sequence recognition built on commonly used linear models where sequence preference is quantified by a matrix of nucleotide energies, the PWM. We consider a collective mode of nucleotide recognition by Klf4, parameterized with a single additional parameter (the coupling constant $J$). This Ising model of sequence-dependent Klf4 binding captures the measured binding energies with sub-$k_BT$ accuracy. Our Ising model contrasts with more complex data-driven approaches, such as hidden Markov models and deep-learning models, in that it yields an accurate sequence-dependent binding energy for sequences that are close to and far away from the reference sequence.

We conclude that *in vitro* binding energies measured with a library of short DNA oligos, via the use of an Ising model, captures Klf4 binding patterns on individual long DNA molecules stretched in the optical tweezer and predict occupancy statistics of Klf4 along the entire genome in human cells. While it is well established that the energetics of TF-DNA interaction are crucial to sequence targeting by TFs, we do not conclude that TF-binding solely obeys equilibrium statistics (Zoller et al., 2022). For example, we expect that non-equilibrium processes that impact TF-DNA binding in a sequence-independent manner affect the statistics of TF binding in a manner that is akin to an increase of an effective temperature. We note, however, that the statistics of genomic Klf4 association is best captured when considering an effective temperature close to the physiological one (Fig. S11). A possibility is that the partaking active processes operate on timescales slower than those of Klf4 binding and unbinding (Hübner and Spector, 2010).

A central determinant of the binding landscape is GC content. The statistical properties of the occupancy patterns (Fig. 5) can indeed be reproduced by an energy model that depends solely on GC content (Fig. S9 F). However, such a simplified model cannot account for the high sequence specificity of Klf4 as evident from bulk competition assay experiments. This is because motif features are lost when evaluating GC content only (Fig. S9 D). On the other hand, the PWM that captures core aspects of sequence-dependent association in the vicinity of motif sequences cannot capture low affinity binding (Fig. 3B, S10). Our Ising model successfully predicts Klf4 occupancy both to low as well as to high affinity sequences, thereby providing a unified quantitative description across the full sequence space.

We have here focused on the association of Klf4 to DNA in a dilute regime, where Klf4-Klf4 interactions do not contribute. Indeed, the occupancy statistics shown in Fig. 5A indicate that, in the cells studied there, Klf4 predominantly resides in a dilute binding regime across most of the genome. Its concentration is therefore likely below the concentration required to trigger a prewetting transition (Morin et al., 2022). An important next step will be to identify cell types and genomic regions where a prewetting threshold might be locally exceeded, enabling droplet formation and the emergence of transcriptional condensates, as has been reported for Klf4 and related factors in living cells (Sharma et al., 2021).

In conclusion, our results favour a picture in which TF–DNA interactions inside cells are governed by a rugged, sequence-dependent energy landscape that, besides individual interactions, also supports emergent, self-organized behaviors such as clustering and condensate formation. In this view, diverse nuclear processes are not imposed externally but arise naturally from the interplay of TF binding and sequence heterogeneity.

# Methods

## A. Expression and purification of Klf4

A codon-optimized Klf4 gene (Uniprot ID (O43474-3), flanked by NotI and AscI restriction sites, was cloned into the pOCC315 FlexiBAC vector. The construct encodes an N-terminal maltose-binding protein (MBP) tag followed by a short linker sequence (SSGR) and a human rhinovirus 3C protease cleavage site. At the C terminus, a linker sequence (GSAGSAAGSG) was inserted upstream of monomeric GFP. MBP-3C-Klf4-GFP (plasmid TH1528) was expressed in $Sf^{+}$ insect cells using the baculovirus system (Lemaitre et al., 2019). Infected cells were incubated for 72 h at 27 °C. Cells were harvested by centrifugation at 4 °C and lysed in ice-cold buffer (50 mM bis-tris propane pH 9.0, 500 mM KCl, 500 mM arginine-HCl, 10 µM $ZnCl_2$, 2 mM $MgCl_2$, 1 mM DTT, 10% glycerol, 0.25 U $ml^{-1}$ benzonase, 1× EDTA-free Protease Inhibitor Cocktail, Selleckchem) using an LM20 Microfluidizer at 5000 psi. All steps were performed at 4 °C. Lysates were clarified by centrifugation (29000 rpm, 1 h, 4 °C) and filtered (0.2 µm PES).

The supernatant was incubated with amylose resin (NEB) for 1 h at 4 °C, washed, and transferred to Econo-Pac gravity columns (Bio-Rad) for further washing with wash buffer II (50 mM bis-tris propane pH 9.0, 1 M KCl, 500 mM arginine-HCl, 10 µM $ZnCl_2$, 1mM DTT, 10% glycerol) and wash buffer I (50 mM bis-tris propane pH 9.0, 500mM KCl, 500 mM arginine-HCl, 10 µM $ZnCl_2$, 1mM DTT, 10% glycerol). Protein was eluted with buffer containing 10 mM maltose. The MBP tag was cleaved overnight at 4 °C using in-house 3C protease.

Cleaved protein was concentrated (Vivaspin 50 kDa MWCO) and purified by size-exclusion chromatography (Superdex 200, GE Healthcare) in buffer (50 mM bis-tris propane pH 9.0, 500 mM KCl, 5% glycerol, 1 mM DTT). Peak fractions were pooled and concentrated. Protein concentration was determined by absorbance at $A_{280}$ and $A_{488}$, the $A_{260}/A_{280}$ ratio was monitored to ensure nucleic acid contamination was less than $0.6$. Protein quality was assessed by mass photometry, analytical SEC, and nanoDSF. Purified Klf4-GFP was used for FA assays within one week and stored at $4°$C.

## B. DNA Oligonucleotides

Unlabeled oligos were ordered as double-stranded and lyophilized, they were purified via HPLC to eliminate remaining single strand DNA. Oligos were stored at $-20°C$ until use. Prior to use, oligos were dissolved in water, and concentrations were measured manually using a NanoDrop spectrophotometer (IMPLEN) at $A_{260}$. Concentrations were calculated based on the extinction coefficient of each double-stranded oligo. A 1 mM stock solution was prepared in water and used to create a 10-fold serial dilution with the ECHO dispensing system.

The labeled reference oligo $S_{\text{ref}*}$ was ordered as double-stranded with an ATTO550 fluorophore attached to the 3' end (Table S1) and lyophilized. Following annealing, it was purified via HPLC to eliminate single strands. The labeled oligo was dissolved in water, and its concentration was measured at $A_{260}$ and $A_{550}$ using a NanoDrop spectrophotometer (IMPLEN). The concentration was calculated considering the correction factor of 0.23 for ATTO550. The labeled oligo was aliquoted, stored at $-20°C$, and used only once per experiment. All competition experiments were conducted using the same batch of labeled oligo. All measured oligos are 17 bp double-stranded duplexes (Table S1). They consist of a $5'$ `GGAG` segment, a variable 8-bp central region, and a $3'$ `GGCTG` segment. All DNA oligos used in present study were purchased from Ella Biotech (Fürstenfeldbruck, Germany).

## C. Competition FA assay

Competition fluorescence anisotropy (FA) assays were performed in a total reaction volume of 50 µl using assay buffer (25 mM Hepes, pH 7.4, 125 mM KCl, 1 mM DTT, 0.1 mg $ml^{-1}$ BSA). All components and buffers were maintained at 4°C throughout the procedure. Assays were set up in 384-well ULA-coated microplates (PerkinElmer). Unlabeled test double-stranded DNA (dsDNA) oligos (17 bp) were dispensed in a 10-step dilution series (0, 0.05, 0.1, 0.2, 0.4, 0.75, 1.5, 3, 6, and 48 µM) using an ECHO 550 acoustic dispensing system (Labcyte). Klf4-GFP was buffer-exchanged into exchange buffer (25 mM Hepes, pH 7.4, 500 mM KCl, 1 mM DTT, 0.1 mg $ml^{-1}$ BSA) using spin desalting columns (Thermo Scientific), ultracentrifuged at $7.5 \cdot 10^4$ rpm for 30 min, filtered, and quantified by absorbance at $A_{280}$ and $A_{488}$. The protein was adjusted in assay buffer to a final concentration of 250 nM. ATTO550-labeled reference dsDNA was prepared at a final concentration of 50 nM in assay buffer. All components were combined and added to the pre-dispensed unlabeled oligo dilution series in the 384-well plates, followed by centrifugation at 1200 $g$ for 5 min. Measurements were performed using a Tecan SPARK 20M multimode microplate reader (Tecan Group Ltd., Männedorf, Switzerland) set to 21.5°C. Excitation was at 550 nm (10 nm bandwidth) and emission at 625 nm (30 nm bandwidth), using a 560 nm dichroic filter. Plates were shaken at 300 rpm for the first 5 min, and anisotropy was subsequently measured every 5 min over a time course of 3.5 h.

## D. FA assay microplate workflow

An ECHO 550 source list was generated in KNIME, defining dispensing positions for all oligos. Based on this list, oligos were transferred into 384-well microplates using the ECHO 550 acoustic liquid handler. Subsequently, the protein solution and reaction components were added to each well under the established assay conditions. Reaction plates were then measured on the SPARK 20M microplate reader. The output files from the SPARK 20M were imported into KNIME, where raw readout values were mapped to the predefined well positions from the original ECHO 550 source list. Each DNA oligo was measured at least three times.

## E. Protein titration FA assay

Protein titration FA assays were performed similarly to the competition assay. The ATTO550-labeled reference dsDNA oligo $S_{\mathrm{ref}*}$ was used at final concentration of 50 nM in assay buffer. Klf4 was prepared in assay buffer and serially diluted two-fold. The protein dilutions were mixed with labeled DNA in 384-well plates, and anisotropy measurements were conducted under the same conditions as for the competition assay.

## F. Biolayer Interferometry (BLI)

BLI measurements were performed on a GatorPlus system (GatorBio, Palo Alto, CA, USA). Biotinylated reference oligo $S_{\mathrm{ref}}$ (150 nM) was immobilized on streptavidin-coated sensors by loading for 420 s. Sensors were then exposed to a two-fold serial dilution of Klf4-GFP, starting at 82 nM, to record association (4200 s) and dissociation (300 s) kinetics. Experiments were conducted in 25 mM HEPES (pH 7.4), 125 mM KCl, 0.05% Tween-20 buffer, with shaking at 400 rpm and data acquisition at 10 Hz. Control assays confirmed no nonspecific binding to biotinylated sensors. Binding kinetics were analyzed using the Gator analysis software (version 2.17.7.0416).

## G. Mass Photometry

Mass photometry experiments were conducted using a TwoMP instrument (Refeyn, Oxford, UK). Klf4-GFP protein was buffer exchanged into exchange buffer as previously described and diluted to 100 nM in the same buffer. For measurement, the sample was further diluted 1:10 directly on the mass photometer by mixing 2 µL of protein with 18 µL of exchange buffer, resulting in a final concentration of 10 nM. Data processing and analysis were performed using the Discover software. Representative measurements are displayed as a histogram of individual mass values. The data was fit by Gaussian distribution.

## H. Size exclusion chromatography coupled to static light scattering

Klf4-GFP molecular weight was determined by size-exclusion chromatography coupled with static light scattering (SEC-SLS) using a Superdex 200 Increase column (Cytiva) connected to an OmniSec GPC max TDA 305 system (Malvern). The system was calibrated in 25 mM HEPES pH 7.4, 1 mM DTT, 500 mM KCl, and 5% glycerol. After calibration with a BSA standard, molecular weight was calculated using static light scattering at 7° and 90° angles, with refractive index detection used to determine protein concentration. Data were processed using OmniSEC software version 4.7.0. Under these conditions, Klf4-GFP was observed as a monomer with an approximate molecular weight of 86 kDa.

## I. Electrophoretic Mobility Shift Assay (EMSA)

EMSA reactions were prepared on ice. Klf4-GFP was buffer-exchanged as described above and used at a final concentration of 1 µM. Reactions were assembled in 25 mM HEPES (pH 7.4), 125 mM KCl, 6% glycerol, 1 mM DTT, 0.1 mg ml$^{-1}$ BSA, and 37.5 ng poly(dI-dC). Cy5-labeled reference DNA (7.5 nM) was added, and samples were incubated for 20 min at 4 °C. Complexes were resolved on pre-run 4-20% Novex TBE gels (Invitrogen) at 250 V for 45 min in 1× TBE at 4 °C. Gels were imaged using a Typhoon FLA 9500 fluorescence scanner (GE Healthcare).

## J. Optical Tweezers Experiments

Single-molecule manipulation experiments were performed on a C-Trap combined optical tweezers and confocal fluorescence microscopy system (LUMICKS, Amsterdam, The Netherlands). Dual-beam optical traps were generated by a 1064-nm infrared laser with independent trap-steering units. Experiments were conducted in a five-channel microfluidic flow cell (LUMICKS). Channels were passivated with 0.1 mg/mL BSA for 1 hour to reduce nonspecific binding. Streptavidin-coated polystyrene beads (4.5 µm, Spherotech) were diluted 1:1000 in buffer prior to use. ATTO647N-labeled $\lambda$-DNA (48.5 kb; LUMICKS) was introduced into the center channel at 0.02 ng/µL, and successful tether formation was verified by characteristic force-extension behavior. All measurements were performed at 28°C.

Before tether formation, Klf4-GFP was flushed into the designated protein channel by applying 0.3-0.4 bar of pressure to the inlet (0.04-inch ID tubing) until the channel was fully exchanged. The pressure was then released, and the protein channel was allowed to equilibrate under no-flow conditions for 2-5 minutes. After establishing a DNA tether in the center channel, the tether was transferred into the pre-equilibrated Klf4-GFP channel and maintained fully immersed in the protein solution. All single-molecule measurements were conducted under static (no-flow) conditions to avoid hydrodynamic perturbations and to allow Klf4-GFP-DNA interactions to reach steady state during data acquisition.

To obtain the steady-state fluorescence profile of the protein along tensioned DNA, bead positions were identified in each image using the protein fluorescence channel, as protein-coated beads exhibit weak autofluorescence. Fluorescence intensities were summed along the y-axis to generate one-dimensional intensity profiles along the x-axis. These profiles were smoothed using a 40-point moving-average filter to reduce noise. Bead centers were determined by fitting Gaussian functions to local intensity maxima. For each peak, a window of ±40 pixels along the x-axis was selected, and the resulting intensity profile was fit using nonlinear least-squares minimization. The center of the fitted Gaussian was taken as the bead position.

The beads have a diameter of 4.34 $\mu$m, due to fluorescence halos from surface-bound protein, regions within 3.5 µm of each bead center are excluded from further analysis. Background fluorescence at steady state is estimated using pixels located more than 3.5 µm from the x-coordinate of either bead center. Pixel values from the final 10 frames (20 s at 0.5 fps) are pooled and fit to a Gaussian distribution, and the mean is taken as the background signal and subtracted from all pixels in every frame. For the same 10 steady-state frames, fluorescence intensities were summed along the x-direction within the region between the bead centers and averaged over frames to obtain a one-dimensional intensity profile along the y-axis. This profile was smoothed using a moving-average filter with a window of 5 pixels to reduce noise. The y-position corresponding to the maximum intensity was taken as the DNA axis. The fluorescence profile is then computed by summing pixel intensities column-wise between the bead centers, restricted to a window extending 5 pixels above and below the DNA axis. The resulting profiles are averaged over time to yield a one-dimensional steady-state fluorescence profile $I(x)$ along the DNA, which is flipped when necessary to ensure consistent molecular orientation.

# Supporting Information for

## *In vitro* binding energies capture Klf4 occupancy across the human genome

**Corresponding Author name.**
**E-mail: hyman@mpi-cbg.de, julicher@pks.mpg.de, grill@mpi-cbg.de**

### This PDF file includes:

Supporting text
Figs. S1 to S11
Table S1

# Supporting Information Text

## Supplementary notes on physical theory

**Analysis of the binding competition assay.** We aim to quantify differences in TF binding energy for a library of DNA sequences. To this end, we analyze fluorescence anisotropy (FA) measurements of a labeled reference oligo in two settings: first without competition (labeled reference oligo only), and then in the presence of competing unlabeled oligos, as in earlier work (Roehrl et al., 2004) and as described in the main text (Fig. 1E,F). In what follows, we first explain how FA measurements allow to quantify TF-DNA in terms of the equilibrium binding constant $K$, when a single species of DNA molecules is present. We then extend this equilibrium description to a competitive binding assay in which an unlabeled oligo competes with a labeled reference oligo for TF binding.

***Quantifying TF-DNA binding with fluorescence anisotropy.*** The measured FA value $F$ of the labeled oligo is captured by a linear relation with anisotropy $F_0$ when no binding occurs and anisotropy $F_1$ when all labeled oligos are bound to a TF:

$$F = F_0 + (F_1 - F_0)P_b \quad . \tag{1}$$

We first consider the simple case of a single oligo at concentration $c = p_1 + s$ and a TF at concentration $p = p_1 + p_0$, where $p_1$ is the concentration of bound TF-oligo complexes, $s$ is the concentration of free (unbound) oligo, and $p_0$ is the concentration of free TF. The fraction of oligos with protein bound is then $P_b = p_1/c$. The binding constant obeys $K = p_0 s/p_1$. Using $K = (p - p_1)(c - p_1)/p_1$, we find

$$P_b = \frac{1 + K/c + p/c}{2}\left(1 - \sqrt{1 - \frac{4p/c}{(1 + K/c + p/c)^2}}\right) \quad . \tag{2}$$

***Competitive binding assay.*** To capture competitive binding, we consider the binding of a TF at total concentration $p$ to two oligos, a labeled oligo at total concentration $c_1$ and an unlabeled oligo at concentration $c_2$. We thus have

$$\begin{aligned} p &= p_0 + p_1 + p_2 \\ c_1 &= p_1 + s_1, \\ c_2 &= p_2 + s_2, \end{aligned} \tag{3}$$

Where $p_0$, $p_1$, and $p_2$ denote the concentrations of free protein, protein bound to the labeled oligo, and protein bound to the unlabeled oligo, respectively. The quantities $s_1$ and $s_2$ are the concentrations of free (=unbound) labeled and unlabeled oligos, respectively. The equilibrium association constants for protein binding to the labeled and unlabeled oligos, denoted $K_1$ and $K_2$ respectively, obey

$$\begin{aligned} K_1 &= \frac{p_0 s_1}{p_1}, \\ K_2 &= \frac{p_0 s_2}{p_2}. \end{aligned} \tag{4}$$

The bound fraction of labeled oligos is given by $P_b = p_1/(p_1 + s_1)$. The binding equilibria can now be written as

$$\begin{aligned} K_1(p - p_0 - p_2) &= p_0(c_1 - p + p_0 + p_2), \\ K_2 p_2 &= p_0(c_2 - p_2) \quad . \end{aligned} \tag{5}$$

Eliminating $p_2$, we obtain a cubic equation for $p_0$

$$p_0^3 + a_2 p_0^2 + a_1 p_0 + a_0 = 0 \tag{6}$$

with

$$a_2 = K_1 + K_2 + c_1 + c_2 - p, \tag{7}$$

$$a_1 = K_2(c_1 - p) + K_1(c_2 - p) + K_1 K_2, \tag{8}$$

$$a_0 = -K_1 K_2 p. \tag{9}$$

Eq. 6 can be solved using trigonometric functions, whereby we obtain

$$P_b = \frac{p_0}{K_1 + p_0}, \tag{10}$$

$$p_0 = \frac{1}{3}\left(2\sqrt{a_2^2 - 3a_1}\,\cos\frac{\Theta}{3} - a_2\right), \tag{11}$$

$$\Theta = \arccos\left[\frac{-2a_2^3 + 9a_2 a_1 - 27a_0}{2\left(a_2^2 - 3a_1\right)^{3/2}}\right]. \tag{12}$$

The fraction of labeled oligo that is protein-bound under given conditions is denoted $P_b(K_1, K_2, p, c_1, c_2)$, which depends on the equilibrium constants, the concentration of active protein and the oligo concentrations For each unlabeled sequence $S$, we express the binding free energy relative to the binding free energy of the reference, given by the labeled oligo

$$\Delta\Delta G_{\mathrm{ref}} = \Delta G - \Delta G_{\mathrm{ref}} = k_B T \ln\left(\frac{K}{K_{\mathrm{ref}}}\right), \qquad [13]$$

where $\Delta G$ and $K$ are the binding free energy and dissociation constant of sequence $S$, and $\Delta G_{\mathrm{ref}}$ and $K_{\mathrm{ref}}$ are the corresponding quantities for the reference sequence $S_{\mathrm{ref}}$. We thus express the bound fraction of TF to the labeled oligo in presence of an oligo of sequence $S$ with total concentration $c$ as $P_b(\Delta\Delta G_{\mathrm{ref}}, K_{\mathrm{ref}}, \alpha p, c_{\mathrm{ref}}, c)$, where $K_{\mathrm{ref}} = K_1$ is the binding affinity of the reference oligo and $K_2 = K$ the affinity of the sequence $S$. The fluorescence anisotropy takes the value $F_0$ for fully unbound labeled oligo and $F_1$ for fully bound labeled oligo. In the experiments, a fraction $\alpha$ of the total protein is considered to be active, so that $1-\alpha$ is the inactive protein fraction and we use $\alpha\, p$ as the relevant TF concentration. The fluorescence anisotropy for sequence $S$ is thus

$$F = F_0 + (F_1 - F_0)\, P_b(\Delta\Delta G_{\mathrm{ref}}, K_{\mathrm{ref}}, \alpha\, p, c_{\mathrm{ref}}, c) \quad . \qquad [14]$$

As the concentrations are known, $K_{\mathrm{ref}}$, $\alpha$, $F_0$, and $F_1$ are determined from a fluorescence anisotropy assay in which unlabeled reference oligo competes out labeled reference oligo (Fig. S2). Thus for each sequence $S$ we have one fit parameter, the relative binding free energy $\Delta\Delta G_{\mathrm{ref}}$.

**Statistical mechanics of sequence-dependent DNA binding by Klf4.** We would like to understand how sequence translates into binding energies. A sequence $S$ of length $N$ is an $N$-tuple $(B_1, .., B_N)$ where $B_i \in \mathcal{B} = \{A, C, G, T\}$ denotes the base at position $i$. The set $\mathcal{B}^N$ of all sequences with length $N$ is a finite space with $4^N$ elements. The sequence-dependent binding energy $\Delta G(S)$ maps points in the $N$-dimensional discrete space to energies in $\mathbb{R}$. As $\mathcal{B}^N$ is finite for any finite $N$, it is in principle possible to specify $\Delta G(S)$ in terms of all the $4^N$ numbers. For $N = 8$, corresponding to the apparent number of nucleotides recognized by Klf4, we need to consider $2^{16} \sim 6.5 \times 10^4$ energies. However, our assay that allows to quantify binding energies with sub-$k_B T$ precision is not feasible at such a scale (even if taking into account the reverse strand cuts this number by half as we discuss further below). Moreover, even if we were able to obtain all those numbers, they would on their own not yield any generalizable understanding of how sequence translates into binding energies. Instead, we aim here to obtain a model for the map $S \to \Delta G$ in terms of a few physically meaningful parameters, while keeping the sub-$k_B T$ precision of our measurements.

***Position-weight matrix and other linear models.*** In the simplest case, a TF is simply sensitive to the number of G/C or A/T base-pairs it binds to. In a linear model, the binding energy is then given by

$$\Delta G = \Delta G_0 + \epsilon_{GC} N_{G/C}, \qquad [15]$$

where $N_{G/C}$ is the number of G/C base pairs in the bound sequence and $\Delta G_0$ is the binding energy of sequences with only A/T base pairs. The sequence-dependence of $\Delta G$ is captured by a single number, $\epsilon_{GC}$, that is the binding energy difference between A/T and G/C base pairs. However, such a simple model is insufficient to capture the sequence-specificity of Klf4 binding (Fig. S9A,D). In fact, replacing a G in the reference motif with a C, thus leaving the GC content unchanged, is often as energetically unfavorable as replacing the G with an A or T (Fig. 3A).

We define the reference motif $S_{\mathrm{motif}}$ of a TF as the DNA sequence the TF binds to most frequently. In the picture of binding energies $\Delta G(S)$, $S_{\mathrm{motif}}$ is the sequence that minimizes $\Delta G(S)$ among sequences with the same length $N$ as $S_{\mathrm{motif}}$. It is insightful to consider the sequence-dependence of $\Delta G(S)$ for sequences that are highly similar to $S_{\mathrm{motif}}$. Sequence similiarity is quantified by the Hamming distance $H$, i.e. the number of positions at which two sequences of length $N$ differ from each other. Sequences are similar to each other, when $H \ll N$. We denote the change in binding energy with respect to the motif sequence as $\Delta\Delta G$. We want to relate $\Delta\Delta G$ to the change in sequence.

To make notation efficient, we first define the base vector $b_i = (A, C, G, T)^T$ to understand the sequence $S = (B_1, ..., B_N)$ in terms of the $N \times 4$ nucleotide matrix $S_{i,j}$.

$$B_i = \sum_{j=1}^{4} S_{i,j} b_j \qquad [16]$$

In words, $S_{i,j} = 1$ iff the nucleotide at position $i$ in $S$ is equal to $b_j$, and $S_{i,j} = 0$ otherwise. For our reference Klf4 motif $S_{\mathrm{motif}}$ translates into a reference nucleotide matrix $S_{i,j}^{\mathrm{motif}}$ as

$$S_{\mathrm{motif}} = \begin{pmatrix} G & G & G & T & G & T & G & G \end{pmatrix}$$

$$S_{i,j}^{\mathrm{motif}} = \begin{pmatrix} 0 & 0 & 0 & 0 & 0 & 0 & 0 & 0 \\ 0 & 0 & 0 & 0 & 0 & 0 & 0 & 0 \\ 1 & 1 & 1 & 0 & 1 & 0 & 1 & 1 \\ 0 & 0 & 0 & 1 & 0 & 1 & 0 & 0 \end{pmatrix}_{i,j} \qquad [17]$$

The change in sequence can then be written as $\Delta S_{i,j} = S_{i,j} - S^{\mathrm{motif}}_{i,j}$. For $H = 1$, $S$ differs from $S_{\mathrm{motif}}$ by only one nucleotide. Thus, $\Delta\Delta G(S)$ can be written in terms of a matrix $E_{i,j}$ that contains all the energy penalties associated with replacing the nucleotide at position $i$ of $S_{\mathrm{motif}}$ with nucleotide $j$:

$$\Delta\Delta G(S) = \sum_{i=1}^{N} \sum_{j=1}^{4} E_{i,j}\, \Delta S_{i,j} \qquad [18]$$

Without loss of generality, we define $E_{i,j} = 0$ for $S^{\mathrm{motif}}_{i,j} = 1$. Then $E_{i,j}$ is uniquely defined by the values $\Delta\Delta G(S)$ for all the sequences $S$ with $H = 1$. Note also that we can then replace $\Delta S_{i,j}$ with $S_{i,j}$ in Eq. 18. This linear model of $\Delta\Delta G(S)$ may then be used as an approximation for sequences with $H > 1$, corresponding to a model where TF-DNA interactions are uncorrelated between different nucleotides.

This is the idea of the position-weight matrix (PWM). From experimental data, e.g. ChIP-seq, one first infers a set of sequences $\Omega_{\mathrm{bound}}$ that are bound by the TF of interest. From this one obtains sequence statistics, in terms of nucleotide probabilities at each position in the bound sequences. Specifically, one calculates the PWM $M_{i,j}$ as

$$M_{i,j} = \log\left[\langle S_{i,j}\rangle_{\mathrm{bound}}\right] - \log p^{\mathrm{ref}}_{j}, \qquad [19]$$

where $\langle\, ..\rangle_{\mathrm{bound}}$ denotes the averages over $\Omega_{\mathrm{bound}}$. $p^{\mathrm{ref}}_{j}$ denotes the average nucleotide probability, corresponding to the GC content, of the experimental set of sequences $\Omega$ the TF can interact with.

Under the assumption that nucleotide probabilities are uncorrelated. we can calculate the probability that a sequence $S$ is TF-bound as

$$p_{\mathrm{bound}}(S) = \exp\left[\sum_{i,j} M_{i,j} S_{i,j}\right] / \mathcal{N}. \qquad [20]$$

, where $\mathcal{N}$ is a normalization factor. This corresponds to a purely statistical model of binding probabilities, that is however closely related to the linear energy model (Eq. 18). Specifically $M_{i,j}$ can be mapped to the energy matrix $E_{i,j}$ under the following assumptions: 1) TF binding as determined in the experiments reflects the equilibrium distribution of TF binding, i.e. the binding probability is proportional to the Boltzmann weight $\exp(-\Delta G/k_B T)$. 2) $\Delta\Delta G(S)$ can be approximated by the linear model (Eq. 18). 3) Nucleotide probabilities are uncorrelated in $\Omega$. Under these three assumptions, one finds

$$M_{i,j} = -E_{i,j} + E_0, \qquad [21]$$

where $E_0$ is some reference energy. Furthermore, this equality requires that the $M_{i,j}$ calculated from experimental data, does indeed reflect true binding probabilities. However, methods like ChIP-seq or SELEX typically yield only the probability that a TF binds within some sequence interval that is significantly larger than the typical length $N$ of the bound sequence (e.g. for Klf4: $N = 8$ whereas ChIP-seq yields intervals of lengths $\geq 100$). To calculate the PWM, typically only the sequences that show a high similarity to the reference motif are taken into account. This means that PWMs reflect the statistics of bound sequences in the vicinity of the reference motif, i.e. $M_{i,j}$ quantifies nucleotide probabilities for sequences with $H \sim 1$. This may explain why we find quantitative agreement between $M_{i,j}$ obtained from ChIP-seq data and the $E_{i,j}$ obtained from direct energy measurements of $H = 1$ sequences (Fig. 2A, 5B).

We can generalize these linear single nucleotide models to higher orders to take into account correlations between nucleotides. Generalizing Eq. 18, we write

$$\Delta\Delta G(S) = \sum_{k=0}^{n} \sum_{i_1=1}^{N-k} \sum_{i_2=i_1+1}^{N-k+1} \cdots \sum_{i_k=i_{k-1}+1}^{N} \sum_{j_1,\ldots,j_k=1}^{4} E^{(k)}_{i_1,\ldots,i_k,j_1,\ldots,j_k} \Delta S_{i_1,j_1} \Delta S_{i_2,j_2} \ldots \Delta S_{i_k,j_k}, \qquad [22]$$

where $n \leq N$ is the order of the expansion. For $n = 1$, we recover Eq. 18 with $E^{(1)}_{i,j} = E_{i,j}$. Without loss of generality, we define the energy matrices such that $E^{(k)}_{\ldots,i_m,\ldots,j_m\ldots} = 0$, when $S^{\mathrm{motif}}_{i_m,j_m} = 0$. Then the matrices $E^{(k)}$ with $k \leq n$ are uniquely defined by the binding energies $\Delta\Delta G(S)$ of all sequences $S$ with $H \leq n$. Vice versa, the expansion up to order $n$ exactly captures the sequence dependence of binding energies for $H \leq n$. For a motif of length $N = 8$, there are $3N + 3^2 N(N-1)/2 = 276$ sequences with $H \leq 2$, whereas our library of high-quality oligos contains only 74 sequences.

***Note on strand symmetry.*** Notably, $\Delta G(S)$ has to obey strand symmetry, i.e. $\Delta G(S)$ has to be equal to $\Delta G(S')$ with $S'$ denoting the reverse complement of $S$. This reflects that $S$ and $S'$ define the same double-stranded DNA. However, as TFs interact with the bases of DNA, a DNA-bound TF has a preferred orientation with respect to the two strands of the DNA double helix. For Klf4, crystal structures indicate that the DNA-binding domain mostly interacts with the G/T-rich strand of the motif. Therefore, the binding energy $\Delta G_{\mathrm{strand}}(S)$ for binding to the strand specified by $S$ is distinct from the energy $\Delta G_{\mathrm{strand}}(S')$ of the reverse complement $S'$. For the linear model considered above, we expect that the binding orientation of a TF remains unchanged for sequences close to the reference motif. Therefore, the matrix $E_{i,j}$ (or its higher order generalization $E^{(k)}$) does not obey strand symmetry, i.e. $E_{i,j} \neq E_{N+1-i,5-j}$. This is also supported by the experimental observation that replacing one of the Gs in the reference motif of Klf4 with the complementary C yields an energy penalty that is as large or larger than replacing the G/C base pair with an A/T base pair (Fig. 3A). When considering overall binding to a DNA molecule, both strands have to be taken into account by summing up the Boltzmann weights of the two binding orientation:

$$\Delta G(S) = -\beta^{-1} \log \left( e^{-\beta \Delta G_{\mathrm{strand}}(S)} + e^{-\beta \Delta G_{\mathrm{strand}}(S')} \right), \quad [23]$$

where $\beta = 1/(k_B T)$. As such $\Delta G(S)$ does obey strand symmetry, while $\Delta G_{\mathrm{strand}}$ does not. Notably, when approximating $\Delta G_{\mathrm{strand}}$ with the linear energy model (Eq. 18), $\Delta G_{\mathrm{strand}}(S)$ is linear in $\Delta S_{i,j}$ whereas $\Delta G(S)$ is not. We thus observe that summing over different binding configurations, corresponding to a thermodynamic partition function, allows to capture non-linearities in $\Delta G(S)$ with a linear energy model. This is the rationale behind the Ising model of sequence recognition.

***Ising model of sequence recognition.*** In the following we develop a generic physical model of DNA binding by a TF to capture nucleotide correlations in the sequence preference of a TF, corresponding to non-linearities in $\Delta\Delta G(S)$ when understood as a function of $\Delta S_{i,j}$. We consider a model where a TF can interact with DNA in two distinct binding modes: 1) a highly sequence-dependent binding mode, likely mediated by direct TF-base interactions, enabled by the TF structure that is optimized for a certain sequence motif 2) a more flexible and transient binding mode with weak sequence dependence, likely mediated by electrostatic interactions between TF and DNA-backbone. In contrast to previous studies, we resolve these states at the level of single nucleotides in the bound sequence. This means that we capture the configurational micro-state of the TF-DNA complex in terms of $N$ classical spins $\sigma_i \in \{+1, -1\}$ where $i$ denotes the nucleotide position within the bound sequence. $\sigma_i = +1$ denotes the highly sequence-dependent binding mode, which we term *nucleotide binding*. $\sigma_i = -1$ denotes the weakly sequence-dependent binding mode which we term *alternative binding*.

We consider a thermodynamic model, where the overall binding energy of a TF to a sequence $S$ reflects the sum of statistical weights over all configurational micro-states of the TF-DNA complex. To each micro-state $\boldsymbol{\sigma} = (\sigma_1, ...\sigma_N)$, we assign an energy given by the Hamiltonian $\mathcal{H}(\boldsymbol{\sigma})$. $\mathcal{H}$ defines the sequence-dependence of binding energies $\Delta G$. In order to capture nucleotide-nucleotide correlations while keeping the simplicity of the linear energy model, we consider configurational couplings, i.e. $\mathcal{H}$ is nonlinear in $\boldsymbol{\sigma}$ but linear in terms of $S_{i,j}$. For simplicity, we consider only nearest neigbour couplings. This yields the following Ising Hamiltonian:

$$\mathcal{H}[\boldsymbol{\sigma}] = E_0 + \sum_{i=1}^{N} \left[ -J_i \sigma_i \sigma_{i+1} + \frac{\sigma_i + 1}{2} \Delta\epsilon_i \right] \quad [24]$$

Here $E_0$ is a sequence-independent reference energy corresponding to a chemical potential. $J_i$ is the coupling constant. For simplicity, we consider the coupling to be sequence- and position-independent, i.e. $J_i = J$ for $i \in \{1, ..., N-1\}$ and $J_N = 0$. $\Delta\epsilon_i$ denotes the energy gap between the local configurational states $\sigma_i = \pm 1$. For $\Delta\epsilon_i < 0$, the sequence-dependent *nucleotide binding* mode ($\sigma_i = +1$) is energetically favorable and thus more likely than the *alternative binding* mode ($\sigma_i < -1$). For simplicity, we consider a single nucleotide model for $\Delta\epsilon_i$:

$$\Delta\epsilon_i = \sum_{j=1}^{4} \Delta\epsilon_{i,j} S_{i,j}. \quad [25]$$

Furthermore, we neglect any sequence-dependence in the *alternative binding* mode ($\sigma_i = -1$), meaning that $E_0$ is sequence-independent.

In order to compute thermodynamic quantities such as the overall free energy $\Delta G$ of the bound state, we calculate the partition function as

$$\mathcal{Z} = \left( \prod_{i}^{N} \sum_{\sigma_i \in \{\pm 1\}} \right) e^{-\beta \mathcal{H}} = e^{-\beta E_0} \mathrm{Tr} \left[ \prod_{i=1}^{N} \mathcal{T}_i \right]. \quad [26]$$

Here Tr denotes the trace of the matrix that is the product of transfer matrices $\mathcal{T}_i$ defined as

$$\mathcal{T}_i = \begin{pmatrix} \exp\left(\beta J_i - \beta \frac{\Delta\epsilon_i + \Delta\epsilon_{i+1}}{2}\right) & \exp\left(-\beta J_i - \beta \frac{\Delta\epsilon_i}{2}\right) \\ \exp\left(-\beta J_i - \beta \frac{\Delta\epsilon_{i+1}}{2}\right) & \exp\left(\beta J_i\right) \end{pmatrix}. \quad [27]$$

Then we obtain the strand-specific binding free energy as

$$\Delta G_{\mathrm{strand}}(S) = -k_B T \log \mathcal{Z}(S) = E_0 - k_B T \log \mathcal{Z}_0(S), \quad [28]$$

where $\mathcal{Z}_0 = \mathcal{Z}\big|_{E_0=0}$. Note that $\mathcal{Z}$ is a highly non-linear function in $\Delta S_{i,j}$, even though we consider a single nucleotide model for $\Delta\epsilon_i$. This is because the the sequence-dependence of $\Delta\epsilon_i$ results not only in sequence-dependent energies of *nucleotide binding* but also in a sequence-dependent probability of *nucleotide binding*. We define the probability of *nucleotide binding* ($\sigma_i = +1$) as

$$p(\textit{nucleotide bound}) = \frac{1}{2} + \sum_{i=1}^{N} \frac{\langle \sigma_i \rangle}{2N} \quad [29]$$

where we calculate the average configuration $\langle \sigma_i \rangle$ as

$$\langle \sigma_i \rangle = \mathrm{Tr}\left[ \left( \prod_{j=1}^{i-1} \mathcal{T}_j \right) \begin{pmatrix} 1 & 0 \\ 0 & -1 \end{pmatrix} \left( \prod_{j=i-1}^{N} \mathcal{T}_j \right) \right] \qquad [30]$$

In order to calculate the binding energy for a DNA molecule with sequence $S_{\mathrm{tot}}$ we sum over both strands and all subsequences of length $N$ within $S_{\mathrm{tot}}$:

$$\Delta G(S_{\mathrm{tot}}) = -k_B T \log \left\{ \sum_{i=1}^{N_{\mathrm{tot}}-N} \mathcal{Z}\left[ S_i^{i+N-1} \right] + \mathcal{Z}\left[ (S_i^{i+N-1})' \right] \right\} \qquad [31]$$

Here, $S_i^j$ denotes the subsequence of $S_{\mathrm{tot}}$ that is composed of the nucleotides at positions $i$ through $j$. As discussed in the main text, this model allows us to capture the measured binding energies. To this end, we fit $\Delta\epsilon_i$ and $J$ as we describe in the Supplementary Methods.

## Supplementary Methods

**Analysis of the FA competition assay.** For each concentration of unlabeled oligo, we computed a single FA value by averaging the FA time course from 50 to 180 min. We determined per-sequence competition energies from repeated FA competition measurements while propagating calibration-parameter uncertainty (Figs. S3-S8). Each library sequence was measured at least $n = 3$ times; we denote replicate measurements by $j \in \{1, \ldots, n\}$ and illustrate the procedure for $n = 3$ with $j \in \{1, 2, 3\}$. For each replicate $j$, we performed an inner bootstrap with $n_{\mathrm{rep}} = 6000$ draws to propagate uncertainty in the calibration parameters. In inner draw $r$, we sampled $(F_1, K_1, \alpha)$ with replacement from the global calibration bootstrap table and sampled $F_0$ independently with replacement from the $F_0$ bootstrap distribution. Holding the sampled $(F_0, F_1, K_1, \alpha)$ fixed, we refit the competition energy parameter $\Delta\Delta G_{\mathrm{ref}}$ by using Eq. 14, yielding one replicate value $\Delta\Delta G_{\mathrm{ref},j,r}$. The $n_{\mathrm{rep}}$ values $\{\Delta\Delta G_{\mathrm{ref},j,r}\}_{r=1}^{n_{\mathrm{rep}}}$ define a replicate-specific distribution for each measurement $j$.

To combine repeated measurements into a single per-sequence estimate, we performed a hierarchical (outer) bootstrap with $B_{\mathrm{hier}} = 10{,}000$ draws. In each outer draw $b$, we resampled $n = 3$ replicate indices with replacement from $\{1, 2, 3\}$ (e.g., $\{1, 1, 2\}$). Given the selected indices $\{j_1, j_2, j_3\}$, we drew one value from each corresponding inner-bootstrap distribution, $\Delta\Delta G_{\mathrm{ref},b}^{k} \sim \{\Delta\Delta G_{\mathrm{ref},j_k,r}\}_{r=1}^{n_{\mathrm{rep}}}$ for $k \in \{1, 2, 3\}$, and formed the combined value as

$$\Delta\Delta G_{\mathrm{ref},b} = \frac{1}{3} \sum_{k=1}^{3} \Delta\Delta G_{\mathrm{ref},b}^{k}. \qquad [32]$$

The $B_{\mathrm{hier}}$ values $\{\Delta\Delta G_{\mathrm{ref},b}\}_{b=1}^{B_{\mathrm{hier}}}$ define the combined per-sequence bootstrap distribution. We report the per-sequence $\Delta\Delta G_{\mathrm{ref}}$ as the mean of this distribution and a 95% confidence interval (insets of Figs. S3-S8).

**Fitting of Ising model.** We obtain the parameters of the Ising model as a non-linear fit from the measured binding energies. To this end we perform a parameter scan in the coupling constant $J$ for range of 18 logarithmically spaced values of $\log J/(k_B T)$ between $-2$ and $+2$ (Fig. S10 G-I). For each value of $J$, we determine the nucleotide energies $\Delta\epsilon_i(b_i)$ as a non-linear fit using the MATLAB function `lsqnonlin`, minimizing the RMSD between model $\Delta\Delta G$ and measured $\Delta\Delta G_{\mathrm{ref}}$ of all library sequences with $H \leq 3$ (to this end also the reference energy $E_0$ is determined as the average difference between model and measured relative binding energies). $\Delta\Delta G$ is calculated as in Eq. 31 taking into account both strands of the 17bp sequences of the oligos with a binding window of $N = 8$. Crucially, we constrained the parameters $\Delta\epsilon_i(b_i)$ from above by a maximum nucleotide binding energy $\Delta\epsilon_{\mathrm{max}}$. This constraint is necessary, because values beyond $\sim 10 k_B T$ are ill-defined as further increases in $\Delta\epsilon_i$ leave the binding energy unchanged, since at this point the statistical weight of the alternative binding state is orders of magnitudes larger than the nucleotide binding state. As an initial guess for the nucleotide energies, we use the energy matrix from measurements of $H = 1$ sequences (see Fig. 3A), i.e. we initialize the non-linear fitting with $\Delta\epsilon_i(b) = \Delta\Delta G_{H=1}(b, i) - \max_b \Delta\Delta G_{H=1}(b, i) + \Delta\epsilon_{\mathrm{max}}$, where $\Delta\Delta G_{H=1}(b, i)$ denotes the measured relative binding energy of the $H = 1$ sequence where the base at position $i$ in the reference motif was substituted with base $b$. Thereby, the most unfavorable substitution yields a nucleotide binding energy equal to $\Delta\epsilon_{\mathrm{max}}$. In order to avoid overfitting, we define the nucleotide energy of the most unfavorable substitution to be equal to $\Delta\epsilon_{\mathrm{max}}$ and optimize only the remaining 24 $\Delta\epsilon_i(b_i)$. This way, we need to fit as many parameters as for the energy matrix $E_{i,j}$ of the linear model. For each $J$, we test a range of $\Delta\epsilon_{\mathrm{max}}$ (see Fig. S10G-I). We found that for $J > 1 k_B T$ and $2 k_B T < \Delta\epsilon_{\mathrm{max}} \leq 8 k_B T$, the Ising model allows to accurately capture the measured binding energies of the $H \leq 3$ sequences to which the $\Delta\epsilon_i$ were optimized (RMSD$<0.5 k_B T$). For $J \sim 2 k_B T$ the model yields also quantitatively accurate energy predictions for $H > 3$ (RMSD$\sim 0.8 k_B T$). In the main text, we use $J = 2.1 k_B T$ and $\Delta\epsilon_{\mathrm{max}} = 7.5 k_B T$ for which the maximum of fitting error and prediction error is minimal (Fig. S10 I). We note, however, that fitting and prediction error differ only slightly from this optimum, as long as $J > 1 k_B T$ and $2 k_B T < \Delta\epsilon_{\mathrm{max}} \leq 8 k_B T$ (Fig. S10 G-I).

**Determining Klf4 occupancy in optical tweezer experiments.** Confocal microscopy quantifies Klf4 occupancy along the $\lambda$-DNA in terms of intensity measurements, specifically the sum $I$ of pixel intensities along a band perpendicular to the DNA (see Methods). As the DNA is stretched, we can map a spatial position $X$ [$\mu$m] along the DNA molecule in the optical tweezer to the sequence position $x$ [bp] on the $\lambda$-sequence:

$$x = N_\lambda \frac{X - X_{\mathrm{bead1}} - r_{\mathrm{bead}}}{X_{\mathrm{bead2}} - X_{\mathrm{bead1}} - 2r_{\mathrm{bead}}} \tag{33}$$

Here $N_\lambda = 48,502\,\mathrm{bp}$ is the length of the $\lambda$-sequence, $X_{\mathrm{bead1/2}}$ are the positions of the bead centers and $r_{\mathrm{bead}} = 2.17\,\mu\mathrm{m}$ is the bead radius. Note that we exclude data points within a radius of $3.5\mu\mathrm{m} > r_{\mathrm{bead}}$ due fluorescence halos around the bead (see Methods). Thus, we assess Klf4 binding to only about 80% of the full $\lambda$-sequence. In order to translate the intensity $I$ to occupancy in terms of molecules per base-pair, we made use of intensity measurements at high protein concentrations ($> 100nM$) where the DNA molecule is saturated with TFs. Crucially, this requires a protein construct lacking the IDR to avoid IDR-mediated protein-protein interactions leading to the formation of a pre-wetting layer. Thus, we obtained a protein construct consisting of the DNA-binding domain of Klf4 as well GFP (details to be published elsewhere). We performed experiments as for full-length Klf4 and quantified the fluorescent intensity on the DNA. As expected, we found that the intensity on DNA saturates at protein concentrations $> 100nM$. We determined the maximum intensity $I_{\max}$ at a bulk protein concentration of $400nM$. Then $I/I_{\max}$ yields an estimate of Klf4 occupancy. In order to estimate the number of molecules per basepair, we considered a maximum occupancy (corresponding to the maximum intensity $I_{\max}$) of one molecule per 10bps. We note that this just an order of magnitude estimate.

In order to compare measured Klf4 occupancies to binding energies predicted by the Ising model, we need to translate binding energies $\Delta G(x)$ to occupancies $p(x)$. Here $\Delta G$ is the calculated binding energy at the 8bp interval at position $x$ in the $\lambda$-sequence and $p(x)$ is the predicted binding probability of Klf4 binding to this 8bp interval. In the dilute regime, we have

$$p = \frac{e^{-\beta \Delta G - \mu}}{e^{-\beta \Delta G - \mu} + 1} \tag{34}$$

where $\mu$ is the chemical potential. To estimate $\mu$, we first calculate the total occupancy $\sum p(x)$ for a range of $\mu$ to determine the chemical potential for which the total predicted occupancy agrees with the observed total occupancy as calculated from the average intensity as $\langle I \rangle / I_{\max} \times N_\lambda / (10\mathrm{bp})$.

In order to compare the predicted occupancy to the measured intensity, we need to take into account the limited resolution of the microscope. To this end, we smoothened the calculated $p(x)$ by convolution with a Gaussian kernel with $\sigma = 1\mathrm{kbp}$.

**Analysis of ChIP-seq data.** We obtained ChIP-seq peaks from the ReMap data base as a .bed file using all data from human cell lines, yielding a total of non-unique 187,685 peaks. We determined the peak position as the average of `thickStart` and `thickEnd`. For each peak, we computed the binding energy of the surrounding 100bp interval in the human genome using Eq. 31. We also calculated the binding energy for overlapping 100bp intervals of the human genome (GRCh38) using a spacing of 10bps of the start position of neighboring intervals. Sequences containing undefined nucleotides were discarded. From the thus obtained binding energy distribution of the 100bp intervals of the human genome, we determined energy percentiles. We then assigned the ChIP-seq peaks to energy percentiles according to the calculated energies. As each energy percentile contains an equal number of sequences, the number $N_{\mathrm{peak}}$ of ChIP-seq peaks assigned to an energy percentile directly yields the Klf4-occupancy of this energy percentile (i.e. the fraction of bound sequences) relative to some unknown baseline level. In Fig. 5 we plot the logarithmic occupancy relative to a genome wide average, i.e. $\log N_{\mathrm{peak}} / \langle \log N_{\mathrm{peak}} \rangle$ where we average over all energy percentiles. The calculated energy $\Delta\Delta G$ of a percentile corresponds there to the genome-wide average of calculated energies within the percentile, relative to the genome-wide average energy $\langle \Delta\Delta G \rangle$. Note that in Fig. 5B, we further shift $\Delta\Delta G$ with respect to the genome-wide average energy to make the regime visible where linear models and Ising model agree with each other.

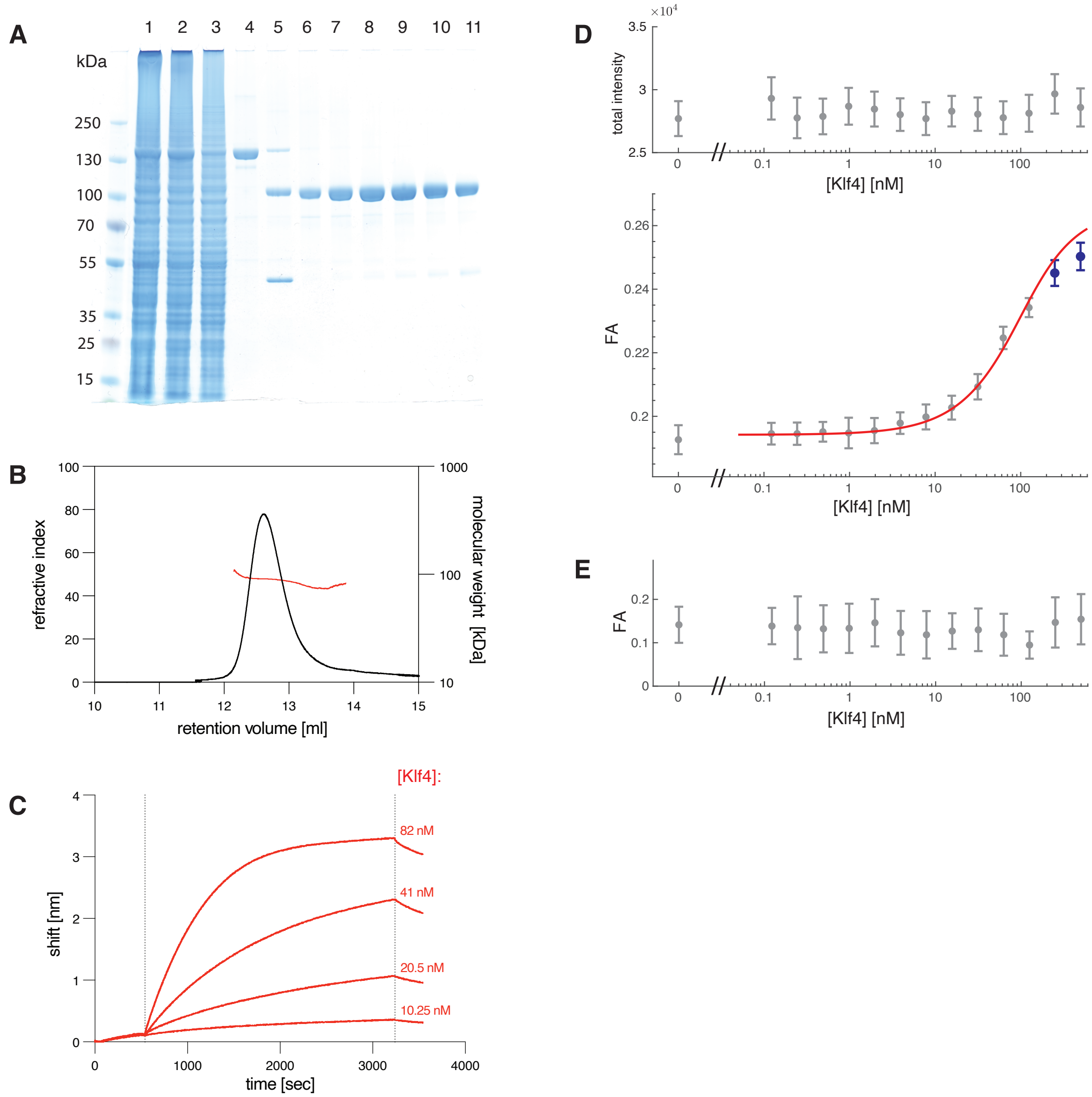

**Fig. S1. Purification and characterization of Klf4-GFP. (A)** Coomassie-stained SDS-PAGE of Klf4-GFP purification: lane 1, total lysate; lane 2, cleared lysate; lane 3, amylose flow-through; lane 4, MBP-Klf4-GFP eluate; lane 5, post-3C protease cleavage releasing Klf4-GFP; lanes 6-11, size-exclusion chromatography (SEC) fractions. MBP was removed during SEC and fractions 7-10 were pooled. **(B)** Size-exclusion chromatography coupled to static light scattering (SLS) of Klf4-GFP. The refractive-index chromatogram (black, left axis) shows a single dominant elution peak. The molar mass derived from the SLS signal (red, right axis) is approximately constant across the peak, indicating a homogeneous monomer Klf4-GFP population with a measured mass of 86 kDa. **(C)** Bio-layer interferometry (BLI) sensograms for Klf4 binding to the reference oligo at the indicated Klf4 concentrations. The association phase begins at the first vertical dotted line, and the dissociation phase begins at the second vertical dotted line. Kinetic parameters were obtained by fitting a 1:1 binding model: $k_{\mathrm{on}} \sim 1.6 \times 10^4$ M$^{-1}$ s$^{-1}$, $k_{\mathrm{off}} \sim 2.4 \times 10^{-4}$ s$^{-1}$, corresponding to $K_d \sim 15$ nM. **(D)** FA titration of Klf4-GFP binding to an ATTO550-labeled reference oligo $S_{\mathrm{ref}*}$. Top, total fluorescence intensity is approximately constant across Klf4 concentrations, indicating no concentration-dependent fluorescence artifacts. Bottom, FA data were fit with Eq. $\mathrm{Eq.}\ (1)$, yielding $K_d \approx 40$ nM; high-concentration points (blue) were excluded from the fit because they exceed the prewetting concentration. Parameters $F_0$, $F_1$, and $\alpha$ were fixed to the values obtained in Fig. S2. **(E)** Control titration without DNA shows no concentration-dependent FA signal in the ATTO550 detection channel, confirming that the GFP on Klf4 does not contribute to the FA readout under these conditions.

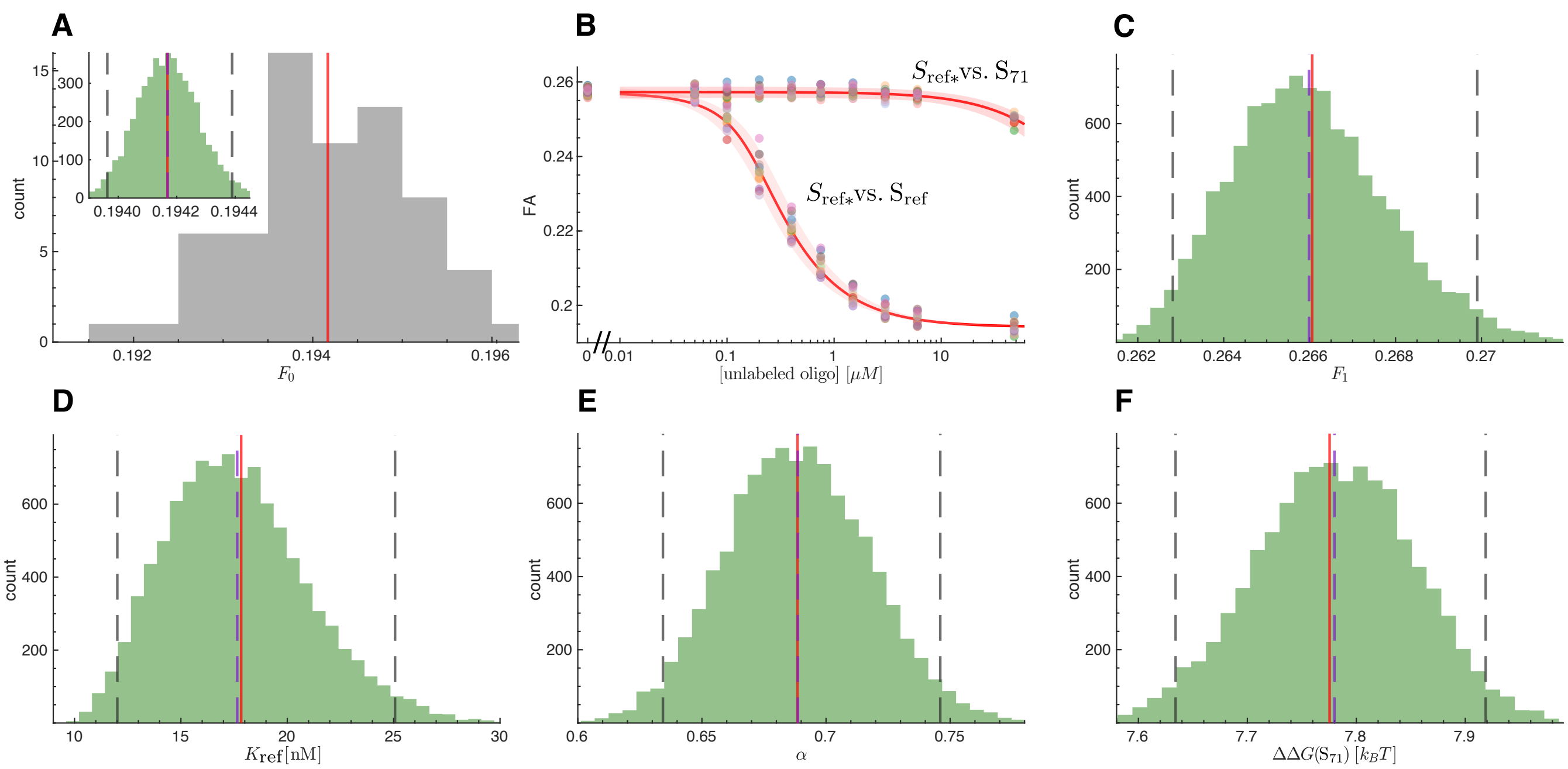

**Fig. S2. FA calibration. (A)** FA of $50\ nM$ ATTO550-labeled $S_{\mathrm{ref}*}$ oligo in the assay condition, approximating the baseline $F_0$. The measured distribution is shown in gray, bin width $5 \cdot 10^{-4}$, $n = 68$. Inset, bootstrap resampling (with replacement) of the $n$ observed measurements was performed for $N = 5000$ replicates, axes as in main panel. The bootstrap sample means are shown as distribution in green, bin width $2 \cdot 10^{-5}$. The magenta vertical dashed line marks the bootstrap mean, and the gray vertical dashed lines indicate the 2.5th and 97.5th percentiles, defining the 95% confidence interval, $F_0 = 0.1942 \pm 0.0002$. Red vertical lines denote the sample mean. **(B)** Plate-wise calibration and global analysis of competition fluorescence anisotropy measurements. Baseline anisotropy $F_0$ was fixed to the value measured in **A**. Each measurement plate included calibration and two competition series in which unlabeled $S_{\mathrm{ref}}$ or $S_{71}$ competed against labeled $S_{\mathrm{ref}*}$ (curves labeled $S_{\mathrm{ref}*}$ vs. $S_{\mathrm{ref}}$ and $S_{\mathrm{ref}*}$ vs. $S_{71}$); each color denotes an independent measurement plate ($n = 11$). Data from each plate were fit with four parameters, $F_1$, $K_{\mathrm{ref}}$, $\alpha$, and $\Delta\Delta G_{\mathrm{ref}}(S_{71})$ using Eq. 14, with protein concentration $p = 250\ nM$, labeled reference oligo concentration $c_{\mathrm{ref}} = 50\ nM$. The red shaded region indicates the envelope of fits across all plates. The red solid line indicates the fit evaluated at the mean parameter values, $\{\langle F_1 \rangle, \langle K_{\mathrm{ref}} \rangle, \langle \alpha \rangle, \langle \Delta\Delta G_{\mathrm{ref}}(S_{71}) \rangle\}$. **(C-F)** Bootstrap distributions of the calibration parameters $F_1$, $K_{\mathrm{ref}}$, $\alpha$, and $\Delta\Delta G_{\mathrm{ref}}(S_{71})$, generated from the fit analysis in **B**. Gray double-dashed vertical lines indicate the 95% confidence intervals: $F_1 = 0.266\ [0.263,\ 0.270]$, $K_{\mathrm{ref}} = 18\ [12,\ 25]\ nM$, $\alpha = 0.69\ [0.63,\ 0.75]$, $\Delta\Delta G(S_{71}) = 7.78\ [7.63,\ 7.92]\ k_B T$. These calibration parameters were used for error propagation of binding-energy estimates in Figs. S3-S8.

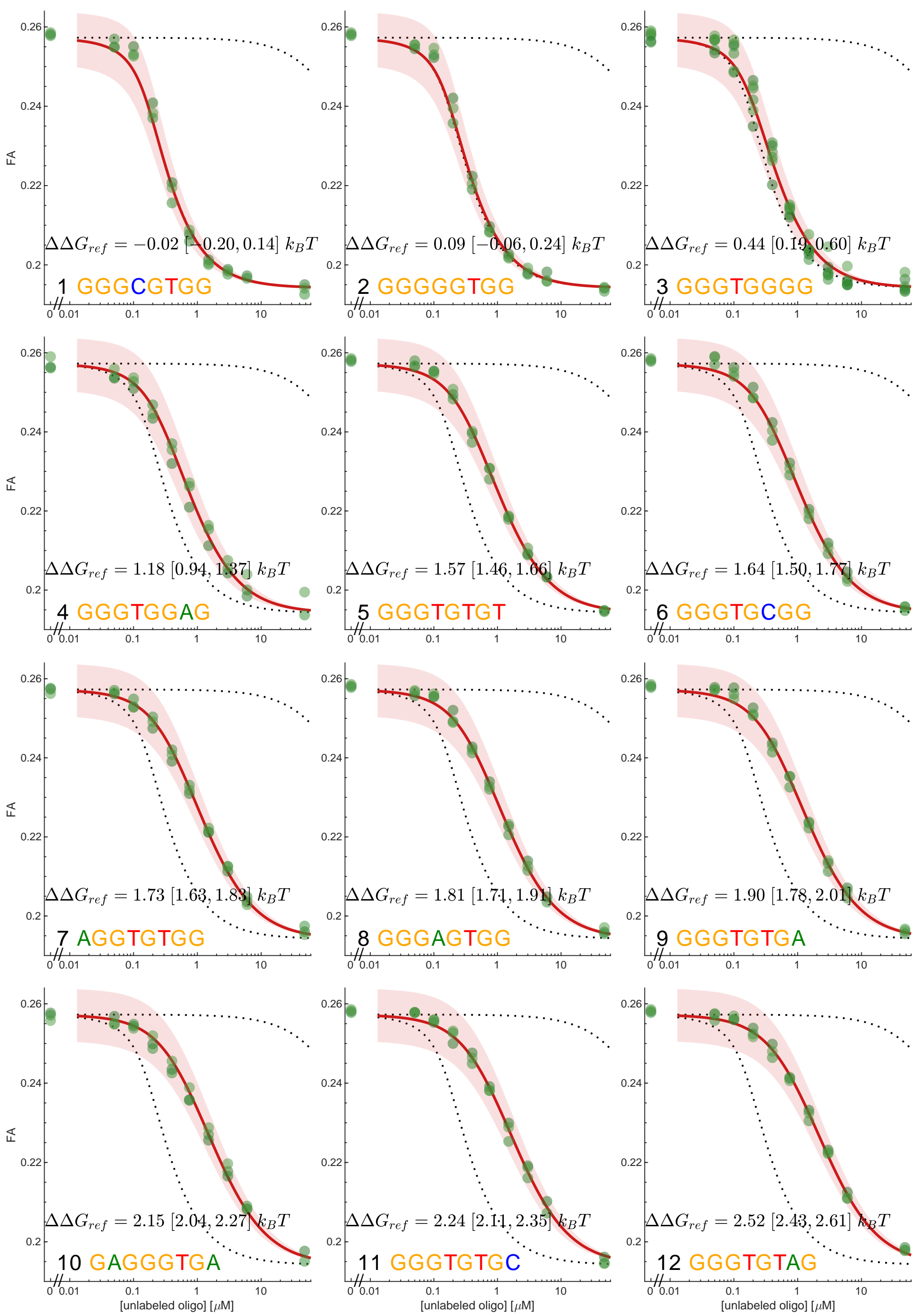

**Fig. S3. FA competition assays for library sequences 1-12.** FA competition curves for library oligos $S_i$ that compete out the fluorescently labeled reference oligo $S_{\mathrm{ref}*}$ from TF binding. Panels are ordered by sequence index $i$ (Fig. 1G), left to right and top to bottom; green symbols are individual measurements (at least three repeats per $S_i$). Each repeat was fit with Eq. 14 using $\Delta\Delta G_{\mathrm{ref}}$ as the only sequence-specific parameter (all other parameters fixed from the global calibration; Fig. S2). The red line shows Eq. 14 evaluated at the mean fitted $\Delta\Delta G_{\mathrm{ref}}$ across repeats; the inset reports this mean with its $95\%$ bootstrap confidence interval (Fig. S2). Black dotted curves show the two reference calibration fits from Fig. S2B (upper and lower), plotted for comparison.

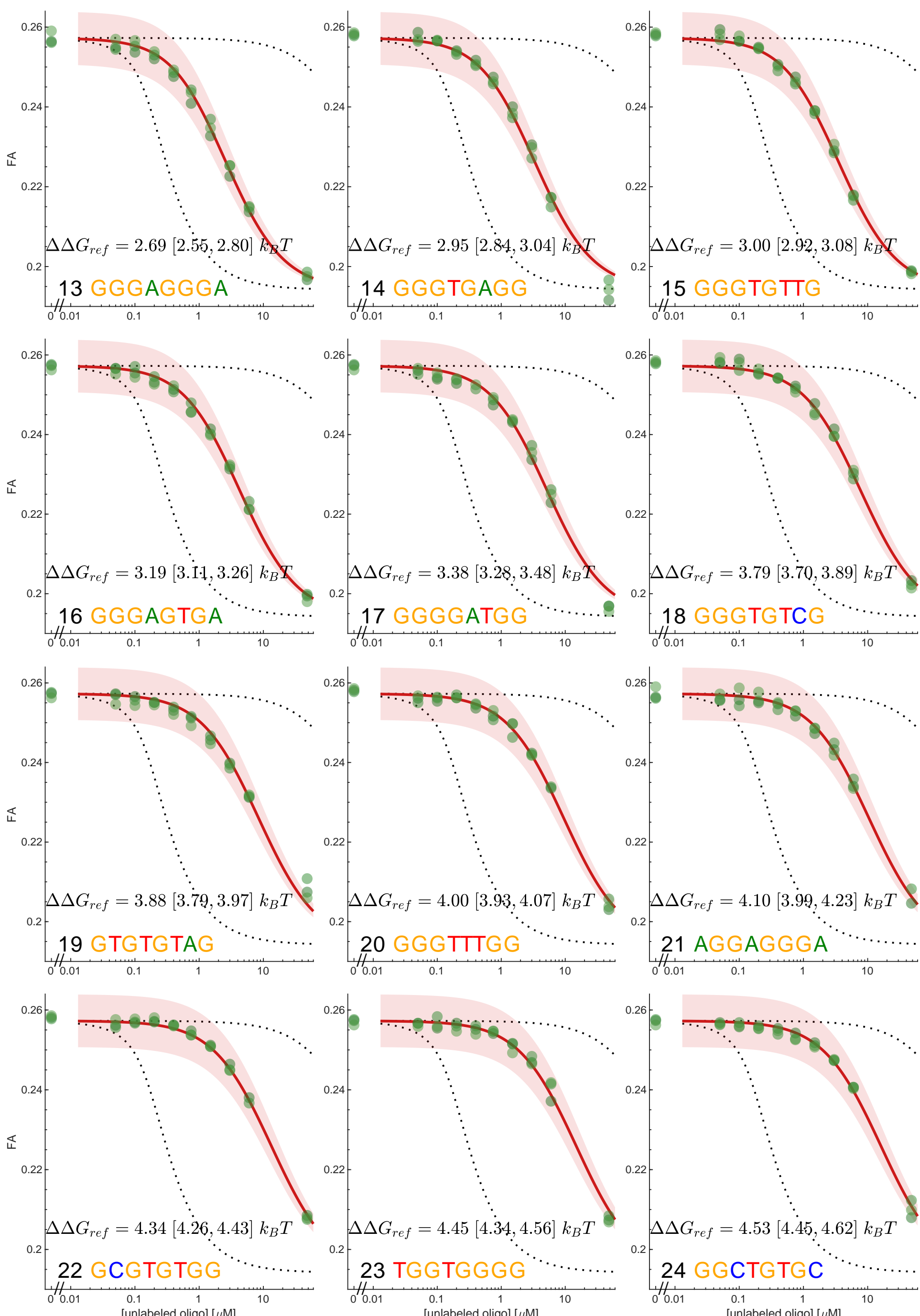

**Fig. S4. FA competition assays for library sequences 13-24.** FA competition curves for library oligos $S_i$ that compete out the fluorescently labeled reference oligo $S_{\mathrm{ref}*}$ from TF binding. Panels are ordered by sequence index $i$ (Fig. 1G), left to right and top to bottom; green symbols are individual measurements (at least three repeats per $S_i$). Each repeat was fit with Eq. 14 using $\Delta\Delta G_{\mathrm{ref}}$ as the only sequence-specific parameter (all other parameters fixed from the global calibration; Fig. S2). The red line shows Eq. 14 evaluated at the mean fitted $\Delta\Delta G_{\mathrm{ref}}$ across repeats; the inset reports this mean with its $95\%$ bootstrap confidence interval (Fig. S2). Black dotted curves show the two reference calibration fits from Fig. S2B (upper and lower), plotted for comparison.

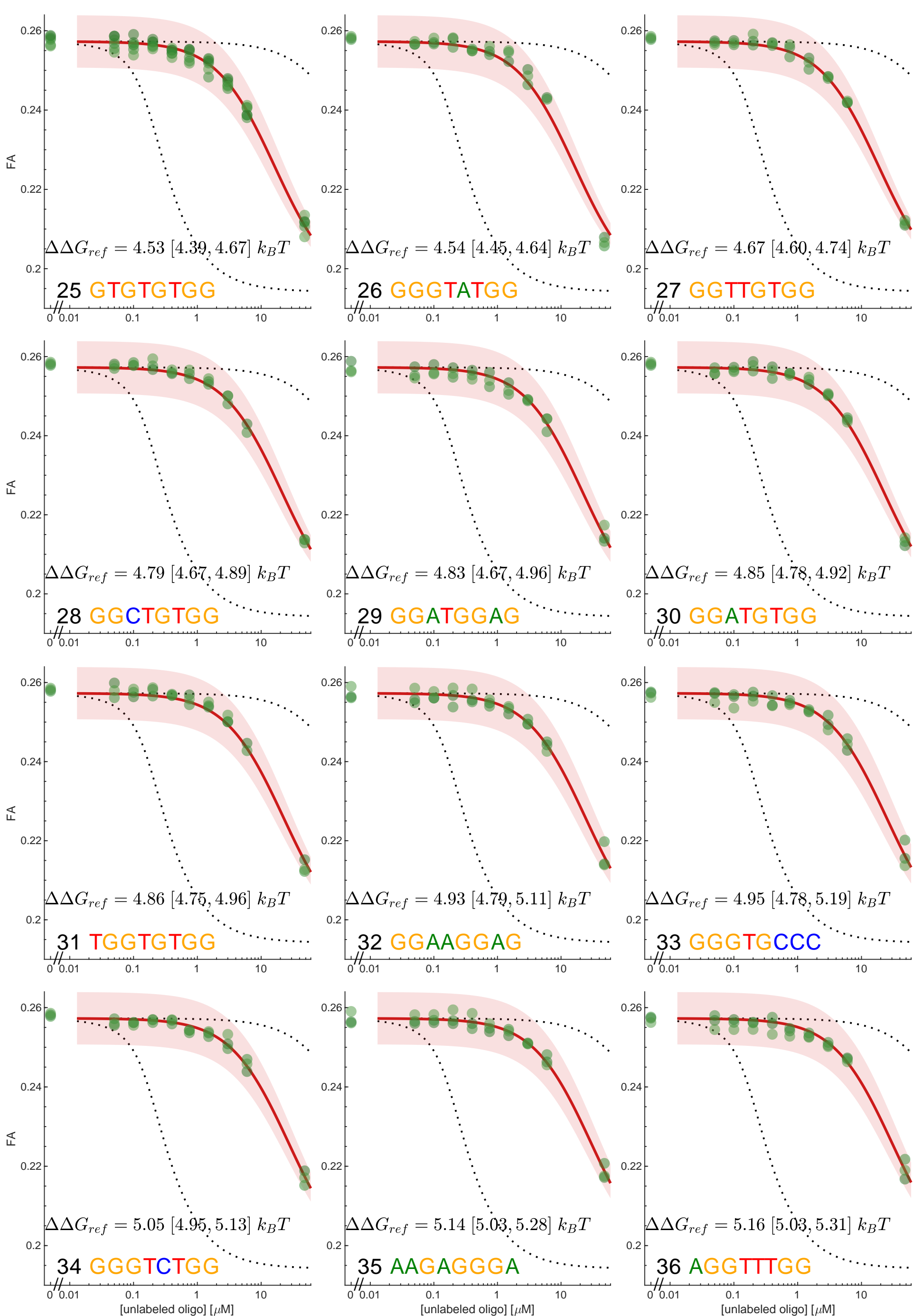

**Fig. S5. FA competition assays for library sequences 25-36.** FA competition curves for library oligos $S_i$ that compete out the fluorescently labeled reference oligo $S_{\mathrm{ref}*}$ from TF binding. Panels are ordered by sequence index $i$ (Fig. 1G), left to right and top to bottom; green symbols are individual measurements (at least three repeats per $S_i$). Each repeat was fit with Eq. 14 using $\Delta\Delta G_{\mathrm{ref}}$ as the only sequence-specific parameter (all other parameters fixed from the global calibration; Fig. S2). The red line shows Eq. 14 evaluated at the mean fitted $\Delta\Delta G_{\mathrm{ref}}$ across repeats; the inset reports this mean with its $95\%$ bootstrap confidence interval (Fig. S2). Black dotted curves show the two reference calibration fits from Fig. S2B (upper and lower), plotted for comparison.

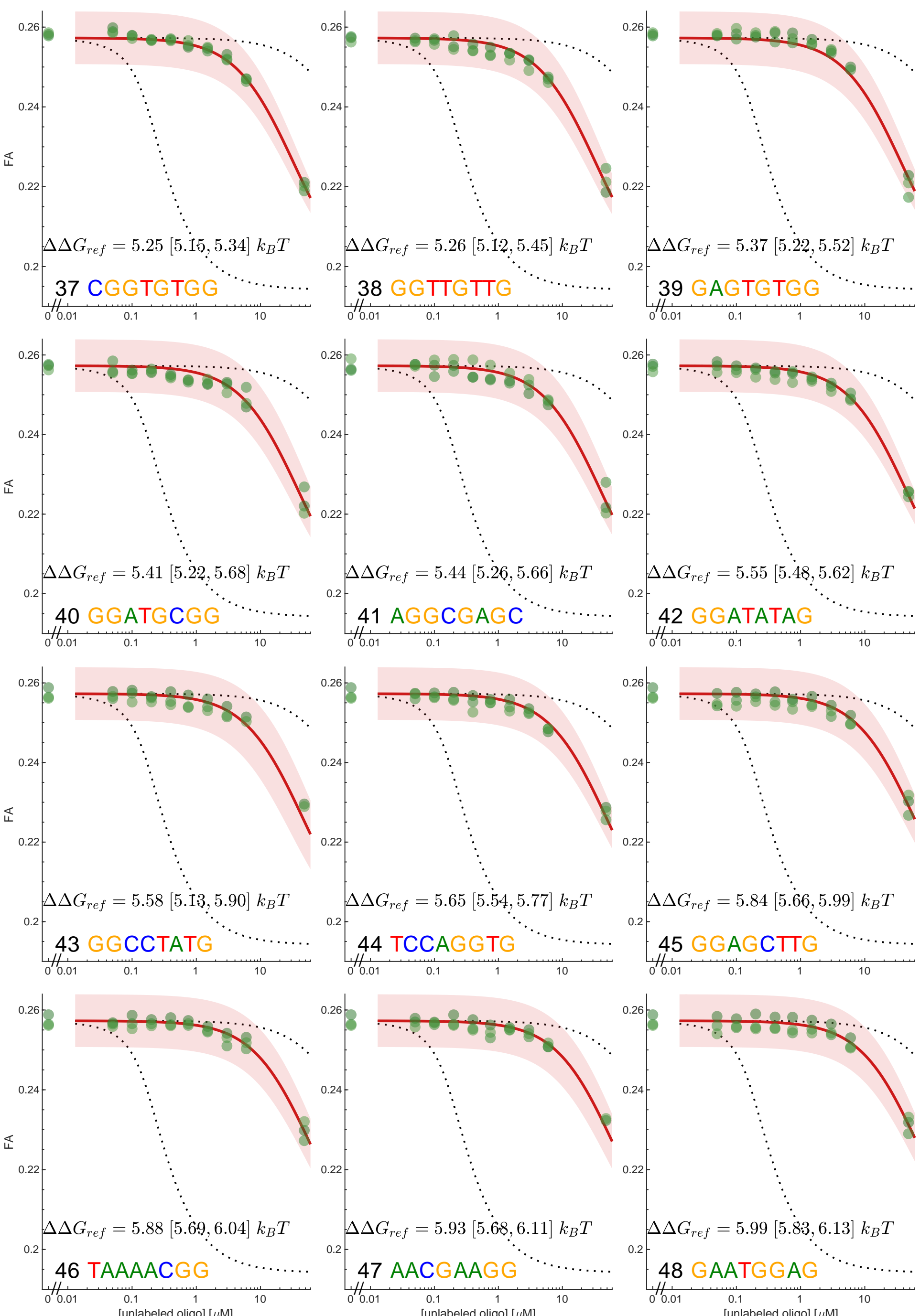

**Fig. S6. FA competition assays for library sequences 37-48.** FA competition curves for library oligos $S_i$ that compete out the fluorescently labeled reference oligo $S_{\mathrm{ref}*}$ from TF binding. Panels are ordered by sequence index $i$ (Fig. 1G), left to right and top to bottom; green symbols are individual measurements (at least three repeats per $S_i$). Each repeat was fit with Eq. 14 using $\Delta\Delta G_{\mathrm{ref}}$ as the only sequence-specific parameter (all other parameters fixed from the global calibration; Fig. S2). The red line shows Eq. 14 evaluated at the mean fitted $\Delta\Delta G_{\mathrm{ref}}$ across repeats; the inset reports this mean with its $95\%$ bootstrap confidence interval (Fig. S2). Black dotted curves show the two reference calibration fits from Fig. S2B (upper and lower), plotted for comparison.

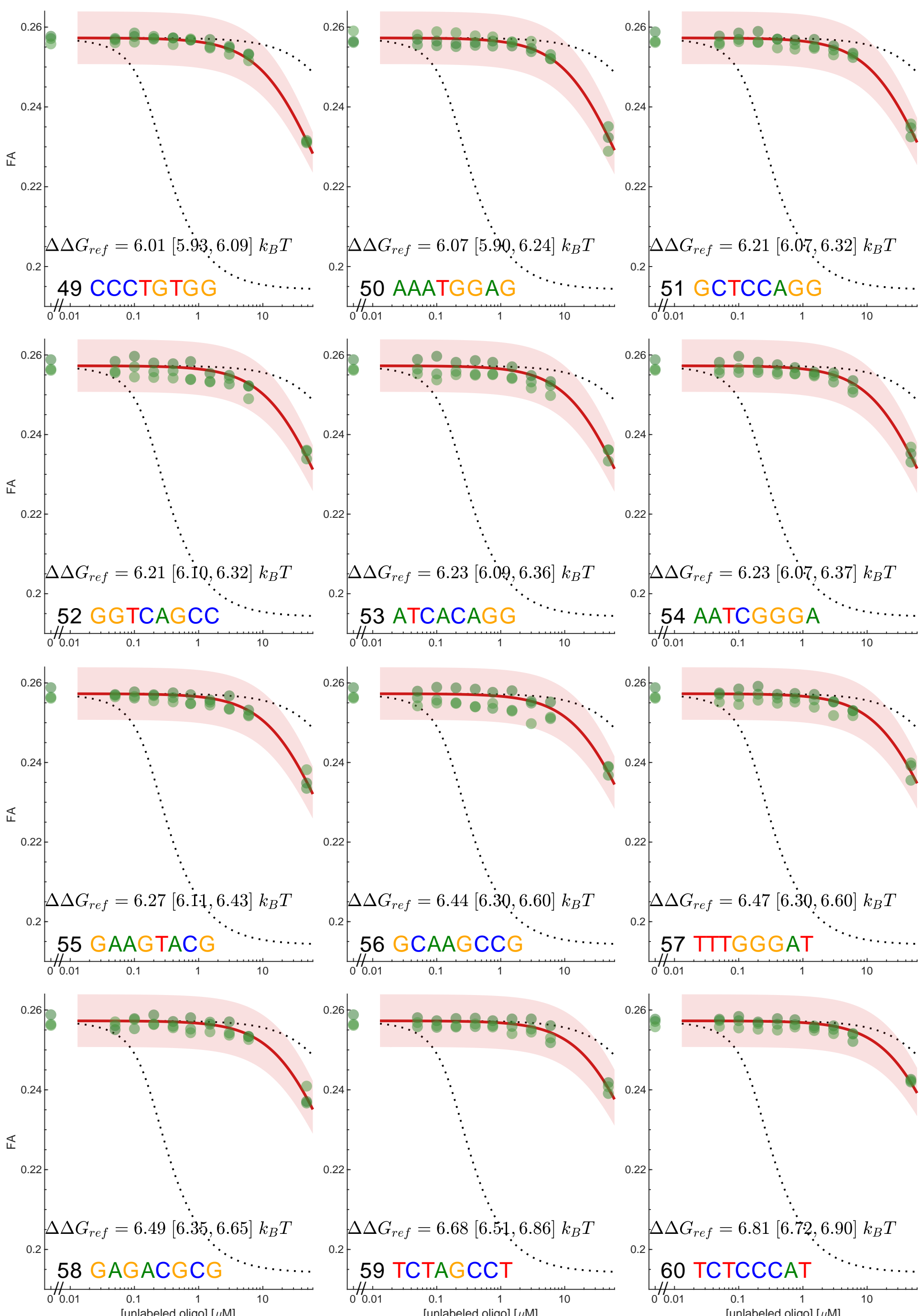

**Fig. S7. FA competition assays for library sequences 49-60.** FA competition curves for library oligos $S_i$ that compete out the fluorescently labeled reference oligo $S_{\mathrm{ref}*}$ from TF binding. Panels are ordered by sequence index $i$ (Fig. 1G), left to right and top to bottom; green symbols are individual measurements (at least three repeats per $S_i$). Each repeat was fit with Eq. 14 using $\Delta\Delta G_{\mathrm{ref}}$ as the only sequence-specific parameter (all other parameters fixed from the global calibration; Fig. S2). The red line shows Eq. 14 evaluated at the mean fitted $\Delta\Delta G_{\mathrm{ref}}$ across repeats; the inset reports this mean with its $95\%$ bootstrap confidence interval (Fig. S2). Black dotted curves show the two reference calibration fits from Fig. S2B (upper and lower), plotted for comparison.

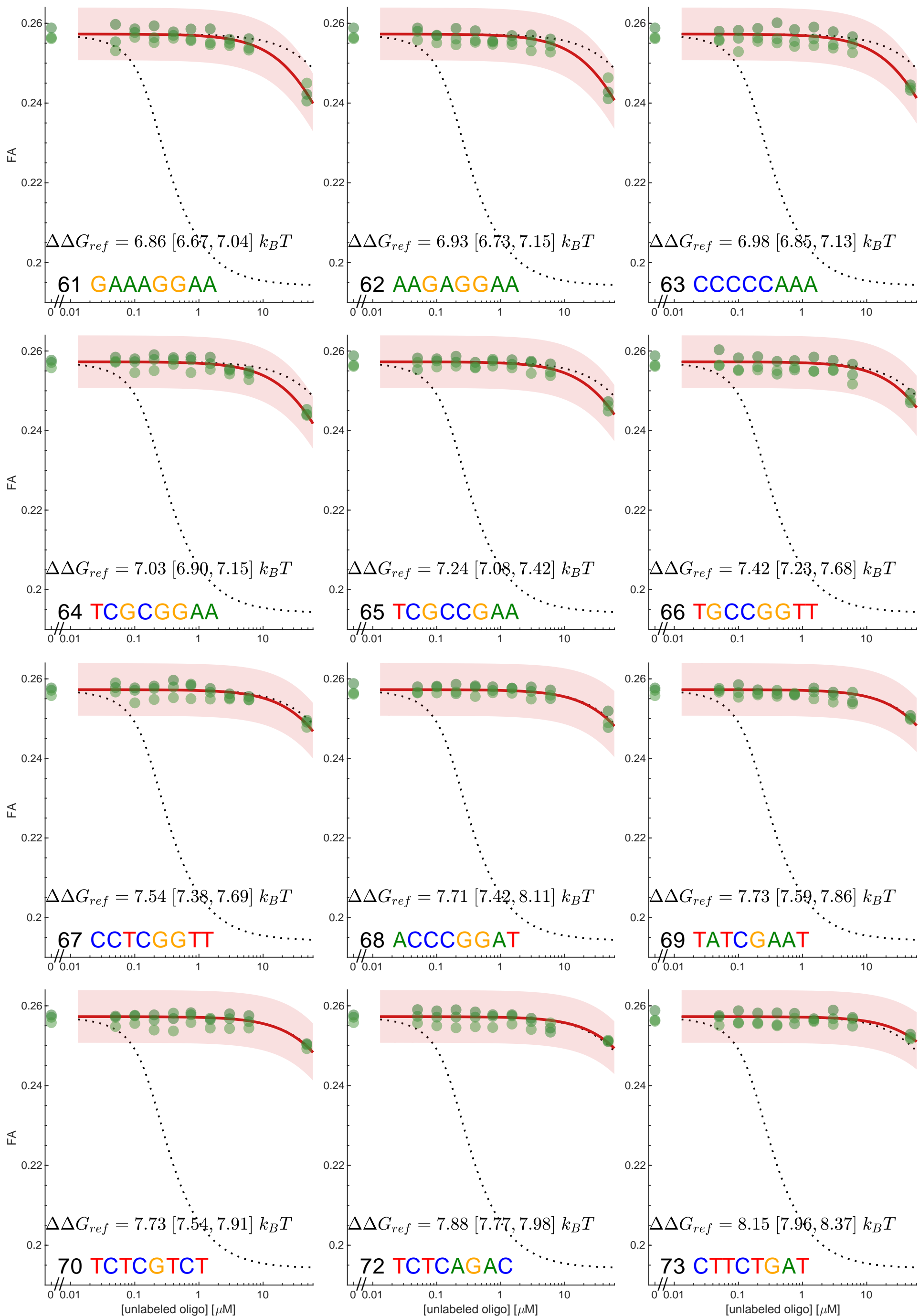

**Fig. S8. FA competition assays for library sequences 61-70,72,73.** FA competition curves for library oligos $S_i$ that compete out the fluorescently labeled reference oligo $S_{\mathrm{ref}*}$ from TF binding. Panels are ordered by sequence index $i$ (Fig. 1G), left to right and top to bottom; green symbols are individual measurements (at least three repeats per $S_i$). Each repeat was fit with Eq. 14 using $\Delta\Delta G_{\mathrm{ref}}$ as the only sequence-specific parameter (all other parameters fixed from the global calibration; Fig. S2). The red line shows Eq. 14 evaluated at the mean fitted $\Delta\Delta G_{\mathrm{ref}}$ across repeats; the inset reports this mean with its $95\%$ bootstrap confidence interval (Fig. S2). Black dotted curves show the two reference calibration fits from Fig. S2B (upper and lower), plotted for comparison.

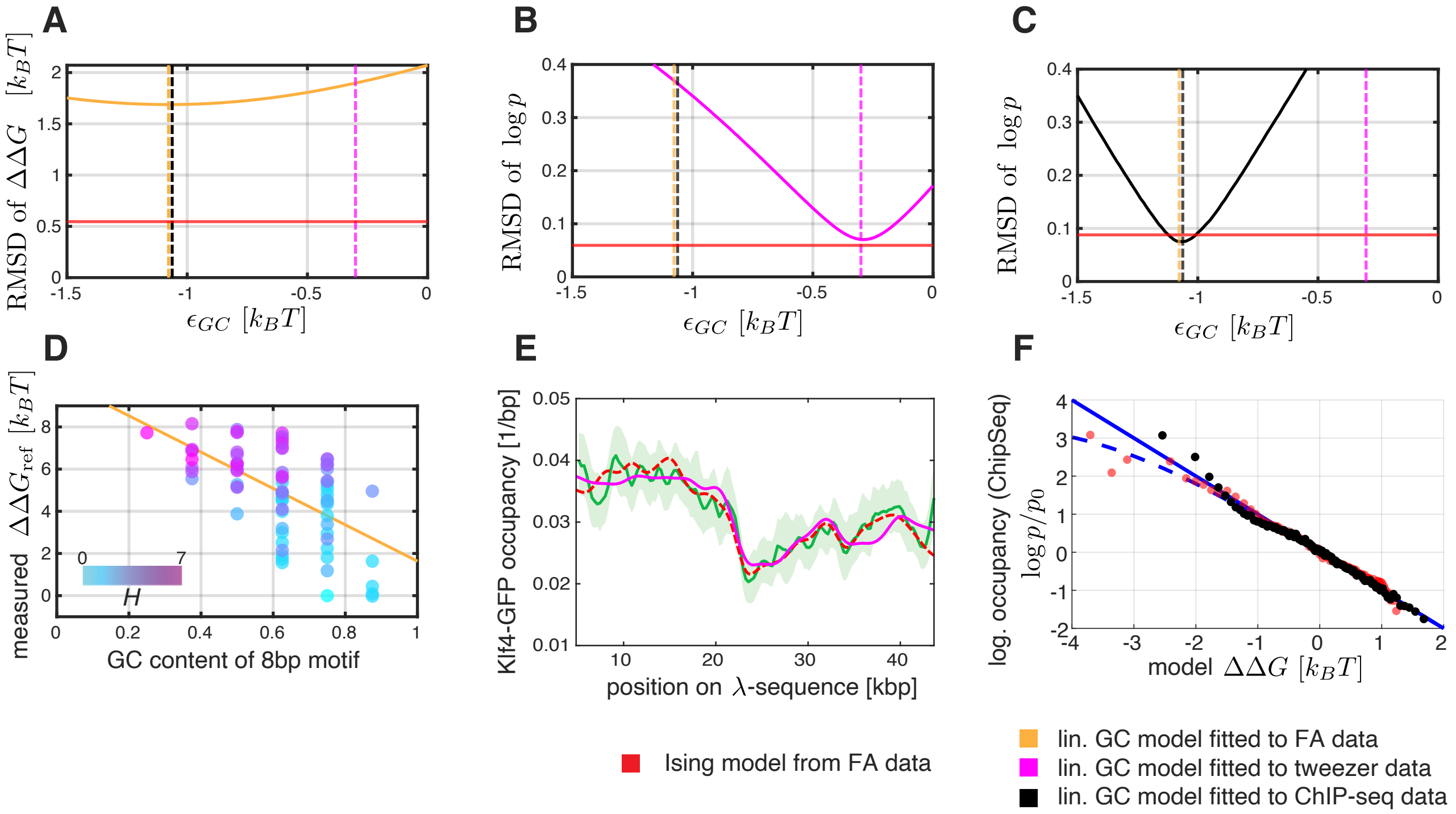

**Fig. S9. Linear GC content models fail to capture Klf4 binding across scales**. **(A)** Fitting error (solid orange curve, root-mean-square deviation (RMSD) of $\Delta\Delta G$) as a function of the G/C dependence $\epsilon_{G/C}$ of binding energy in a linear GC model (Eq. 15). Dashed lines indicate the GC slopes $\epsilon_{G/C}$ that capture best the FA measurements of $\Delta\Delta G$ (orange), the measured Klf4-GFP occupancy in the optical tweezer experiment (magenta) or the Klf4 occupancy inferred from ChIP-seq data (black), respectively. Red solid line indicates the RMSD of $\Delta\Delta G$ across all measured sequences for the Ising model fitted to the measured $\Delta\Delta G$ of sequences with $H \leq 3$ (see Fig. 3). **(B)** Fitting error (solid magneta curve) of the linear GC model as a function of the GC slope $\epsilon_{G/C}$. We calculate here the RMSD of measured logarithmic Klf4-GFP occupancy (see Fig. 4) with respect to the model prediction of the linear GC model (magenta) or the Ising model obtained from FA data (red). We calculate the occupancy from the energies from linear GC model and Ising model analougously. Dashed lines identical to **A**. **(C)** Analogous to **B** but for Klf4 occupancy of 100bp intervals of the human genome as calculated from ChIP-seq data as in Fig. 5. Note that we exclude here the lower 3 percentiles of binding energies to focus on the dilute regime where $\log p = -\Delta\Delta G + \mathrm{const.}$, **(D)** Scatter plot of measured relative binding energies $\Delta\Delta G_{\mathrm{ref}}$ as a function of the GC content of the central 8bp motif (scatter points colored according to Hamming distance as in Fig. 3). Yellow solid line indicates best fit of linear GC model ($\epsilon_{GC} = -1.07 k_BT$). **(E)** Klf4 occupancy along the $\lambda$ sequence as in Fig. 4B (green and red curves are identical between here and Fig. 4B). Solid magenta line shows the model prediction for a linear GC model with $\epsilon_{GC} = -0.3 k_BT$. **(F)** Logarithmic Klf4 occupancy of 100 bp intervals of the human genome as in Fig. 5 (red scatter points and blue curves are identical to Fig. 5A). Black points show Klf4 occupancy (obtained from ChIP-seq) for percentiles in binding energy as calculated from a linear GC content model with $\epsilon_{GC} = -1.1 k_BT$. Note that the GC model does not exhibit the expected saturation at low binding energies (= high affinity sequences).

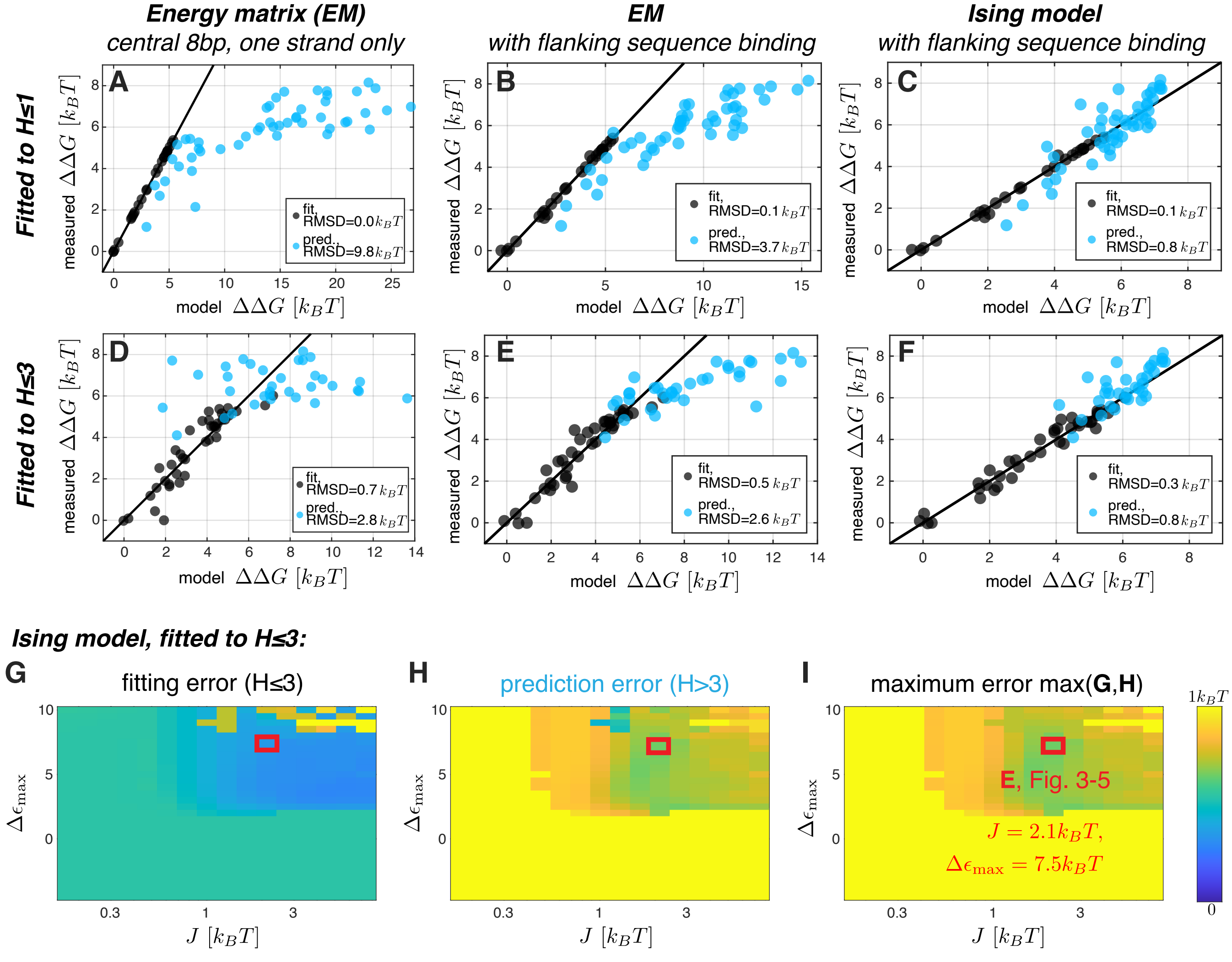

**Fig. S10. Binding energy models inferred from FA measurements (A-F)** Scatter plots of measured binding energies vs. model binding energies, where the models are fitted to the black data points, i.e. the sequences with $H \leq 1$ (**A-C**) and $H \leq 3$ (**D-F**) respectively. Blue data points are the remaining sequences, amounting to a model prediction that is used to validate the model in terms of the root-mean square deviation (RMSD). **(A,D)** Energy matrix (EM) (Eq. 18 with $N = 8$) considering binding only to the central 8bps, neglecting strand symmetry. Data points in **A** are identical to Fig. 3B. Energy matrices were obtained via linear regression. **(B,E)** Statistical mechanics model of TF binding, where we use an energy matrix to predict binding energies to 8bp sequences and sum up the Boltzmann weights of all 8bp intervals of the two strands of the 17bp oligos as a partition function (generalizing Eq. 23 analogous to Eq. 31). Entries of the energy matrix were obtained as a non-linear fit, similar to the Ising model but with no constraints on model parameters. **(C,F)** Ising model of sequence recognition with $J = 2.1k_BT$ and $\Delta\epsilon_{\mathrm{max}} = 7.5k_BT$. Data points in **F** are identical to Fig. 3D. **G-I** Evaluation of fitting error and model predictions of the Ising model for a parameter scan in terms of the coupling constant $J$ and the maximum nucleotide binding energy $\Delta\epsilon_{\mathrm{max}}$. For each parameter pair $J$, $\epsilon_{\mathrm{max}}$ the nucleotide energies $\Delta\epsilon_i(b_i)$ are obtained as a non-linear fit to the the measured energies of the $H \leq 3$ sequences as described in the Suppl. Methods. Color indicates the RMSD between model and measured binding energy with yellow corresponding to a poor agreement with an RMSD$\geq 1k_BT$. Red square indicates the parameter pair used throughout the main text which was selected as the best model in terms of fitting error and model prediction (see **I**). **(G)** fitting error, i.e. RMSD for the energies of $H \leq 3$ sequences. **(H)** prediction error, i.e. RMSD for the energies of the library sequences with $H > 3$. **(I)** maximum of prediction and fitting error, which we use to evaluate the overall accuracy of the model to select the best parameters $J, \Delta\epsilon_{\mathrm{max}}$. We find a good fit and accurate model predictions for $J \sim 2k_BT$ and $2k_BT < \Delta\epsilon_{\mathrm{max}} < 9k_BT$.

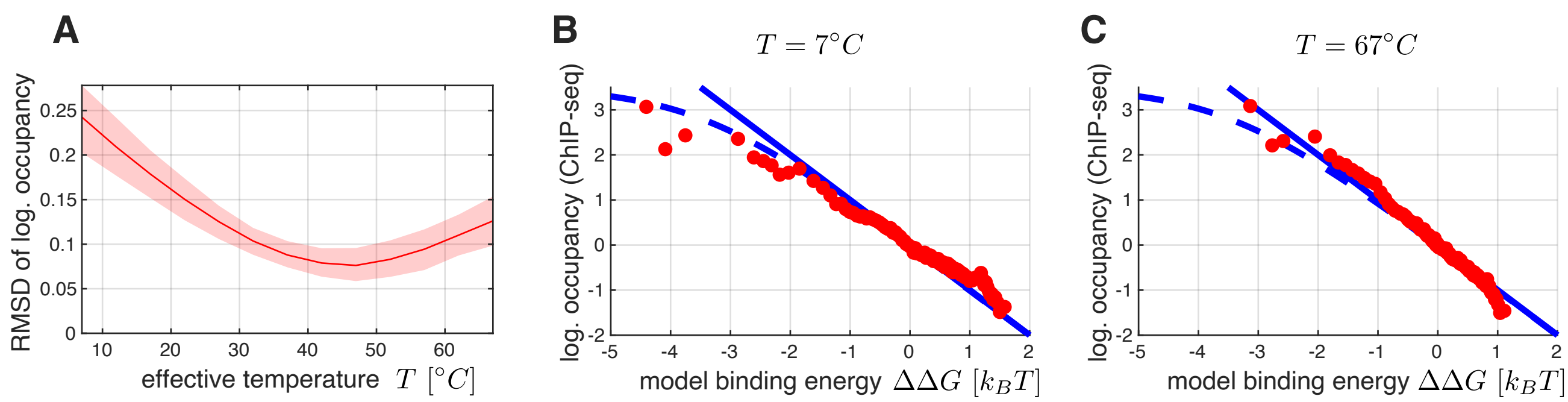

**Fig. S11. Effective temperature of Klf4 binding on the human genome** We compare here ChIP-seq data of Klf4 binding in human cells to predictions of binding energy by the Ising model as in Fig. 5. In contrast to Fig. 5, we consider an effective temperature in the Ising model that can be distinct from the physiological temperature of $T \sim 37^\circ C$. **(A)**To quantify agreement between model prediction and Klf4 occupancy on the human genome, we compute the root-mean-square deviation of the relative apparent occupancy $\log p - \langle \log p \rangle$ with respect to the negative relative binding energy $-\Delta\Delta G + \langle \Delta\Delta G \rangle$ as calculated with our Ising model (Fig. 3) for a range of effective temperatures. $\langle .. \rangle$ denotes here the genome-wide average. We use here the percentiles of 100bp intervals of the human genome as in Fig. 5. Solid line is the RMSD, when excluding the strongest binding lowest 3 percentiles in $\Delta\Delta G$. Shaded area indicates a confidence interval from bootstrapping, treating the genomic percentiles as samples. **(B)** Scatter plot (red) of Klf4 occupancy of energy percentiles of 100bp intervals of the human genome as in Fig. 5A, but for an effective temperature $T = 7^\circ C$. Blue solid line: $y = -x$ corresponding to dilute binding. Blue dashed line; $y = -\log\left(e^{x+\mu} + 1\right) + \mu$ with chemical potential $\mu = 3.5 k_B T$ corresponding to saturated binding, as in Fig. 5. **(C)** analogous to **B** with $T = 67^\circ C$.

**Table S1. Oligos used in present work**

| label | $5' - 3'$ sense | $3' - 5'$ antisense |
|---|---|---|
| ref* | GGAGGGGTGTGGGGCTG-ATTO550 | CAGCCCCACACCCCTCC |
| ref | GGAGGGGTGTGGGGCTG | CAGCCCCACACCCCTCC |
| 1 | GGAGGGGCGTGGGGCTG | CAGCCCCACGCCCCTCC |
| 2 | GGAGGGGGGTGGGGCTG | CAGCCCCACCCCCCTCC |
| 3 | GGAGGGGTGGGGGGCTG | CAGCCCCCCACCCCTCC |
| 4 | GGAGGGGTGGAGGGCTG | CAGCCCTCCACCCCTCC |
| 5 | GGAGGGGTGTGTGGCTG | CAGCCACACACCCCTCC |
| 6 | GGAGGGGTGCGGGGCTG | CAGCCCCGCACCCCTCC |
| 7 | GGAGAGGTGTGGGGCTG | CAGCCCCACACCTCTCC |
| 8 | GGAGGGGAGTGGGGCTG | CAGCCCCACTCCCCTCC |
| 9 | GGAGGGGTGTGAGGCTG | CAGCCTCACACCCCTCC |
| 10 | GGAGGAGGGTGAGGCTG | CAGCCTCACCCTCCTCC |
| 11 | GGAGGGGTGTGCGGCTG | CAGCCGCACACCCCTCC |
| 12 | GGAGGGGTGTAGGGCTG | CAGCCCTACACCCCTCC |
| 13 | GGAGGGGAGGGAGGCTG | CAGCCTCCCTCCCCTCC |
| 14 | GGAGGGGTGAGGGGCTG | CAGCCCCTCACCCCTCC |
| 15 | GGAGGGGTGTTGGGCTG | CAGCCCAACACCCCTCC |
| 16 | GGAGGGGAGTGAGGCTG | CAGCCTCACTCCCCTCC |
| 17 | GGAGGGGGATGGGGCTG | CAGCCCCATCCCCCTCC |
| 18 | GGAGGGGTGTCGGGCTG | CAGCCCGACACCCCTCC |
| 19 | GGAGGTGTGTAGGGCTG | CAGCCCTACACACCTCC |
| 20 | GGAGGGGTTTGGGGCTG | CAGCCCCAAACCCCTCC |
| 21 | GGAGAGGAGGGAGGCTG | CAGCCTCCCTCCTCTCC |
| 22 | GGAGGCGTGTGGGGCTG | CAGCCCCACACGCCTCC |
| 23 | GGAGTGGTGGGGGGCTG | CAGCCCCCCACCACTCC |
| 24 | GGAGGGCTGTGCGGCTG | CAGCCGCACAGCCCTCC |
| 25 | GGAGGTGTGTGGGGCTG | CAGCCCCACACACCTCC |
| 26 | GGAGGGGTATGGGGCTG | CAGCCCCATACCCCTCC |
| 27 | GGAGGGTTGTGGGGCTG | CAGCCCCACAACCCTCC |
| 28 | GGAGGGCTGTGGGGCTG | CAGCCCCACAGCCCTCC |
| 29 | GGAGGGATGGAGGGCTG | CAGCCCTCCATCCCTCC |
| 30 | GGAGGGATGTGGGGCTG | CAGCCCCACATCCCTCC |
| 31 | GGAGTGGTGTGGGGCTG | CAGCCCCACACCACTCC |
| 32 | GGAGGGAAGGAGGGCTG | CAGCCCTCCTTCCCTCC |
| 33 | GGAGGGGTGCCCGGCTG | CAGCCGGGCACCCCTCC |
| 34 | GGAGGGGTCTGGGGCTG | CAGCCCCAGACCCCTCC |
| 35 | GGAGAAGAGGGAGGCTG | CAGCCTCCCTCTTCTCC |
| 36 | GGAGAGGTTTGGGGCTG | CAGCCCCAAACCTCTCC |
| 37 | GGAGCGGTGTGGGGCTG | CAGCCCCACACCGCTCC |
| 38 | GGAGGGTTGTTGGGCTG | CAGCCCAACAACCCTCC |
| 39 | GGAGGAGTGTGGGGCTG | CAGCCCCACACTCCTCC |
| 40 | GGAGGGATGCGGGGCTG | CAGCCCCGCATCCCTCC |
| 41 | GGAGAGGCGAGCGGCTG | CAGCCGCTCGCCTCTCC |
| 42 | GGAGGGATATAGGGCTG | CAGCCCTATATCCCTCC |
| 43 | GGAGGGCCTATGGGCTG | CAGCCCATAGGCCCTCC |
| 44 | GGAGTCCAGGTGGGCTG | CAGCCCACCTGGACTCC |
| 45 | GGAGGGAGCTTGGGCTG | CAGCCCAAGCTCCCTCC |
| 46 | GGAGTAAAACGGGGCTG | CAGCCCCGTTTTACTCC |
| 47 | GGAGAACGAAGGGGCTG | CAGCCCCTTCGTTCTCC |
| 48 | GGAGGAATGGAGGGCTG | CAGCCCTCCATTCCTCC |
| 49 | GGAGCCCTGTGGGGCTG | CAGCCCCACAGGGCTCC |
| 50 | GGAGAAATGGAGGGCTG | CAGCCCTCCATTTCTCC |
| 51 | GGAGGCTCCAGGGGCTG | CAGCCCCTGGAGCCTCC |
| 52 | GGAGGGTCAGCCGGCTG | CAGCCGGCTGACCCTCC |
| 53 | GGAGATCACAGGGGCTG | CAGCCCCTGTGATCTCC |
| 54 | GGAGAATCGGGAGGCTG | CAGCCTCCCGATTCTCC |
| 55 | GGAGGAAGTACGGGCTG | CAGCCCGTACTTCCTCC |
| 56 | GGAGGCAAGCCGGGCTG | CAGCCCGGCTTGCCTCC |
| 57 | GGAGTTTGGGATGGCTG | CAGCCATCCCAAACTCC |
| 58 | GGAGGAGACGCGGGCTG | CAGCCCGCGTCTCCTCC |
| 59 | GGAGTCTAGCCTGGCTG | CAGCCAGGCTAGACTCC |
| 60 | GGAGTCTCCCATGGCTG | CAGCCATGGGAGACTCC |
| 61 | GGAGGAAAGGAAGGCTG | CAGCCTTCCTTTCCTCC |
| 62 | GGAGAAGAGGAAGGCTG | CAGCCTTCCTCTTCTCC |
| 63 | GGAGCCCCCAAAGGCTG | CAGCCTTTGGGGGCTCC |
| 64 | GGAGTCGCGGAAGGCTG | CAGCCTTCCGCGACTCC |
| 65 | GGAGTCGCCGAAGGCTG | CAGCCTTCGGCGACTCC |
| 66 | GGAGTGCCGGTTGGCTG | CAGCCAACCGGCACTCC |
| 67 | GGAGCCTCGGTTGGCTG | CAGCCAACCGAGGCTCC |
| 68 | GGAGACCCGGATGGCTG | CAGCCATCCGGGTCTCC |
| 69 | GGAGTATCGAATGGCTG | CAGCCATTCGATACTCC |
| 70 | GGAGTCTCGTCTGGCTG | CAGCCAGACGAGACTCC |
| 71 | GGAGTCTCGGATGGCTG | CAGCCATCCGAGACTCC |
| 72 | GGAGTCTCAGACGGCTG | CAGCCGTCTGAGACTCC |
| 73 | GGAGCTTCTGATGGCTG | CAGCCATCAGAAGCTCC |